\documentclass[lettersize,journal]{IEEEtran}
\usepackage{filecontents,url}
\usepackage{cite}
\usepackage{amsmath,amssymb,amsfonts,bm}
\usepackage[caption=false,font=normalsize,labelfont=sf,textfont=sf]{subfig}
\usepackage{comment}
\usepackage{xcolor}
\usepackage{soul}
\usepackage{framed}
\colorlet{shadecolor}{blue!20}
\usepackage{xcolor}
\usepackage{graphicx}
\usepackage{multirow}
\usepackage{pdfpages}
\usepackage[linesnumbered,ruled,noend]{algorithm2e}

\SetAlgoCaptionSeparator{:}
\SetNlSty{texttt}{}{:}
\usepackage[T1]{fontenc}
\usepackage[font=small,labelfont=bf,tableposition=top]{caption}

\DeclareCaptionLabelFormat{andtable}{#1~#2  \&  \tablename~\thetable}

\SetAlgoCaptionSeparator{:}
\SetNlSty{texttt}{}{:}

\newcommand{\eg}{\emph{e.g.}\xspace}
\newcommand{\etc}{\emph{etc}\xspace}

\newcommand{\wrt}{ with respect to }

\newtheorem{definition}{Definition}[section]

\def\BibTeX{{\rm B\kern-.05em{\sc i\kern-.025em b}\kern-.08em
    T\kern-.1667em\lower.7ex\hbox{E}\kern-.125em}}
    
\begin{document}



\title{{\sf AutoModel}: Automatic Synthesis of Models from Communication Traces of SoC Designs\\
}

\author{
\IEEEauthorblockN{ Md Rubel Ahmed, Bardia Nadimi, Hao Zheng}
\\University of South Florida, Tampa, FL\\
\{mdrubelahmed, bnadimi, haozheng\}@usf.edu}

\maketitle


\begin{abstract}

Modeling system-level behaviors of intricate System-on-Chip (SoC) designs is crucial for design analysis, testing, and validation. However, the complexity and volume of SoC traces pose significant challenges in this task. This paper proposes an approach to automatically infer concise and abstract models from SoC communication traces, capturing the system-level protocols that govern message exchange and coordination between design blocks for various system functions.
This approach, referred to as \emph{model synthesis}, constructs a causality graph with annotations obtained from the SoC traces. 
The annotated causality graph represents all potential causality relations among messages under consideration.
Next, a constraint satisfaction problem is formulated from the causality graph, which is then solved by a satisfiability modulo theories (SMT) solver to find satisfying solutions.  
Finally, finite state models are extracted from the generated solutions, which can be used to explain and understand the input traces.
The proposed approach is validated through experiments using synthetic traces obtained from simulating a transaction-level model of a multicore SoC design and traces collected from running real programs on a realistic multicore SoC modeled with gem5.

\end{abstract}




\section{Introduction}
\label{motivation}


Modern complex system designs incorporate numerous functional blocks sourced from diverse origins. During runtime, these blocks communicate and coordinate with each other through intricate system-level protocols to execute sophisticated functions. System executions involve a high degree of concurrency, with multiple system-level transactions adhering to these protocols being executed simultaneously. Based on experiences, it has been observed that communication among different blocks is a significant source of design and runtime errors~\cite{cusp, ps}. To ensure thorough verification of the communication behavior in system designs, well-defined and comprehensive protocol specifications are indispensable.
However, in practice, \emph{such specifications are usually unavailable, incomplete, or even contain errors. }
They often become outdated and disconnected from the design implementation as the design progresses. 

Various methods and approaches (\eg~\cite{Yang:2006,Liu:2013,Mrowca:2019:LTS:3316781.3317847,natasa2020}) have been proposed to mine patterns or models from traces in different contexts. However, these methods are inadequate for handling the system communication traces considered in this work. Existing methods extract patterns or models from a trace based on strong temporal dependencies among events identified in the trace. In the case of communication traces resulting from concurrent executions of multiple system transactions, events showing strong temporal dependencies may not be correlated according to the true dependencies in the ground truth specifications. As a consequence, the existing mining methods often produce numerous patterns, many of which are either meaningless or invalid. Moreover, these methods may fail to extract a model encompassing all valid patterns, leading to large, incomprehensible, or even misleading extracted models.

To address the above challenges, this paper describes a method that can automatically infer reduced, concise, and accurate models from the communication traces of complex system designs obtained from simulation or emulation.  
The traces considered in this paper are represented as sequences of messages exchanged among design blocks. 
Given a set of input traces, this method first constructs a causality graph that captures all possible causality relations among messages in the input traces. 
Then, the input traces are mined to annotate the causality graph with frequencies of the causality relations among different messages. 
Next, a constraint satisfaction problem is formulated for the causality graph, and it is solved by an SMT solver to find satisfying solutions from which finite state models are extracted.  
The extracted models characterize the underlying system-level communication protocols that design blocks follow to generate the input traces.
This paper also presents several techniques that assist in filtering out less likely models or improve the accuracy of the inferred models.

This work's key distinguishing feature is its focus on traces generated from concurrent execution of multiple system transactions. The main objective is to infer system-level protocols that govern the executions of these system transactions.
This paper makes the following \textbf{contributions}.
\begin{itemize}
 
    

    \item To the best of our knowledge, the proposed method is the first to automatically and efficiently infer concise and accurate abstract models from communication traces of complex system designs, effectively characterizing the implemented system-level protocols.

    \item By incorporating readily available design information, the method can avoid finding invalid transitions in the extracted models, resulting in reduced and more comprehensible models.
    
    \item The method is complemented with several optimization techniques that effectively filter out less likely models, thereby enhancing the accuracy and quality of the inferred models.

    \item A novel and fast comprehensive method for evaluating the extracted models is introduced.
    
    \item It is efficient and scalable, as demonstrated by realistic SoC traces in the experiments.
\end{itemize}


The organization of this paper is as follows. 
Section~\ref{sec:rel-work} reviews some closely related work in specification mining. 
Section~\ref{sec:background} provides the necessary background for the related concepts and formulates the problem. 
Section~\ref{sec:gap} discusses the limitations of the previous methods in the context considered in this paper and motivates our method. 
Section~\ref{sec:method} and \ref{sec:opt} describe the proposed model synthesis method and the optimization techniques. 
Section~\ref{sec:results} presents the experimental results, and the last section concludes the paper.


\section{Related Work}
\label{sec:rel-work}

Specification mining is a process that involves extracting patterns from diverse artifacts. The model-based approach, \textit{Synoptic} \cite{Beschastnikh:2011:LEI:2025113.2025151}, mines invariants from logs of sequential execution traces with concurrency recorded in partial order. It then generates a Finite State Machine (FSM) that satisfies the mined invariants. Another tool, \emph{Perracotta}~\cite{Yang:2006}, is a software execution log analysis tool that mines temporal API rules. It utilizes a chaining technique to discover long sequential patterns.

The work \textit{BaySpec} \cite{Mrowca:2019:LTS:3316781.3317847} extracts LTL formulas from Bayesian networks trained with software traces. It requires traces to be clearly partitioned with respect to different functions.
For mining hardware traces, the approaches presented in~\cite{Li2010DAC,Hertz:2013:tcad,Danese:2015:vlsi-soc,Danese:2015:date,Danese:2017:dac,mining_specs_ammons2002} mine assertions from either gate-level representations~\cite{Li2010DAC}, or RTL models~\cite{Chang:2010:aspdac,Hertz:2013:tcad,Danese:2015:vlsi-soc,Danese:2015:date,Danese:2017:dac,wencho_dac2010}.
The work presented in~\cite{Liu:2013} focuses on an assertion mining approach using episode mining from simulation traces of transaction-level models. However, these approaches often lack the capability to find longer patterns, limiting their ability to discover complex communication patterns involving multiple components. A recent study~\cite{Ahmed:mine-msg-flow:2021} addresses a similar problem, but it exclusively mines sequential patterns and does not aim to infer the models pursued in this paper.

The work in~\cite{Noc_fabrics} presents a security property validation method that specifically targets the communication fabrics of SoCs. It constructs a Control Flow Graph (CFG) for each Intellectual Property (IP) in the SoC and efficiently explores connections between various CFGs to verify any given security property. By considering system-level interactions, this study highlights the significance of communication fabrics as a potential point of vulnerability. Consequently, understanding the fabric-level communication model becomes crucial for validation activities.


Our work belongs to the category of model synthesis methods, which aims to identify a model from system execution traces such that the resulting model can accurately capture the input traces. Previous research efforts such as~\cite{Heule:2013} have extended deterministic finite automata inference based on the evidence-driven state merging method~\cite{Lang:1998:EDSM}, formulating it as a graph coloring problem and solving it using a Boolean satisfiability solver. Recently, \emph{Trace2Model}~\cite{natasa2020} was introduced, which learns non-deterministic finite automata (NFA) models from software execution traces using C bounded model checking technique. Similar work can also be found in~\cite{Ulyantsev:2011}.

However, the above approaches do not consider the concurrent nature of communication traces in SoC designs, and they heavily rely on temporal dependencies discovered from traces to identify models. As a result, these methods may not capture the true dependencies that our work aims to find.
\section{Background}
\label{sec:background}

\begin{figure}[tb]
\begin{center}
\begin{tabular}{cc}
\begin{minipage}{1.8in}
\footnotesize
\begin{verbatim}
1 (cpu0:cache:rd_req)
2 (cache:cpu0:rd_resp)
3 (cpu1:cache:rd_req)
4 (cache:cpu1:rd_resp)
5 (cache:mem:rd_req)
6 (mem:cache:rd_resp)
\end{verbatim}
\end{minipage}
& 
\begin{minipage}{1.in}
\includegraphics[height=1.1in]{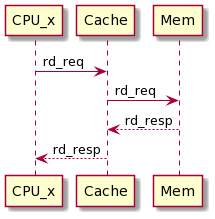}
\end{minipage}
\\
(a) & (b)
\end{tabular}
\vspace{-1pt}
\caption{CPU downstream read flows. (a) Definitions of messages, (b) Message sequence diagram for the flows.  This diagram is parameterized with $x$ which can be $0$ or $1$.}
\label{fig:flow-ex}
\end{center}
\vspace{-15pt}
\end{figure}

System-level protocols are commonly represented as \emph{message flows} in system architecture documents. As depicted in Fig.~\ref{fig:flow-ex}, a message flow concisely describes temporal relations for a set of messages in a system design. For instance, the example in Fig.~\ref{fig:flow-ex} illustrates memory read operations for two CPUs via a shared cache, demonstrating a simple yet representative message flow scenario in a multi-core design.
As shown in Fig.~\ref{fig:flow-ex}(a), each message is a triple $({\tt src: dest: cmd})$ where the ${\tt src}$ denotes the originating component of the message while ${\tt dest}$ denotes the receiving component of the message.  Field ${\tt cmd}$ denotes the operation to be performed at ${\tt dest}$.  
For example, message ${\tt (CPU0: Cache: rd\_req)}$ is the read request from ${\tt CPU0}$ to ${\tt Cache}$.  
Each message flow is associated with a unique \emph{initial} message to indicate initiation of an instance of such flow, and with one or multiple different \emph{terminal} messages to indicate its completion. 
A message flow in a system can have multiple branches, each describing different paths that the system can execute. For instance, in Fig.~\ref{fig:flow-ex}(b), two flows can be instantiated, each with two branches specifying read operations for cache hit or miss scenarios. Taking the flow initiated by ${\tt CPU0}$ as an example, message~1 serves as the initial message of this flow. The two branches of the flow are denoted as $(1,2)$ when message~1 results in a cache hit and $(1,5,6,2)$ when message~1 results in a cache miss. In this context, message indices are utilized to represent the corresponding messages.

During the execution of a system design, multiple instances of flows are executed concurrently. Each flow instance follows a specific path, exchanging messages with runtime information such as memory addresses, transaction IDs, packet IDs \etc. SoC traces typically capture the execution of multiple instances of different flows that occur concurrently.

In this method, an execution \textbf{trace} is 
    $\rho = (\varepsilon_0, \varepsilon_1, \ldots, \varepsilon_n)$
where $\varepsilon_i$ is a set of messages observed at time $i$, and each $m_{i} \in \varepsilon_i$ is an instance of message~$m$ of some flow instance active at time $i$.
The concurrent execution of message flows can happen in two ways: \emph{simultaneously} or in an \emph{interleaved} manner. In the simultaneous case, the size of $\varepsilon_i$ in a trace can be larger than 1, while in the interleaved case, the size of $\varepsilon_i$ in a trace is 1. For the sake of generality, this paper considers traces where the size of $\varepsilon_i$ is larger than or equal to 1. To simplify the presentation, we will use the terms "flows" (or "messages") and "flow instances" (or "message instances") interchangeably when their meanings are clear in the context.

Let us consider the flows depicted in Fig.~\ref{fig:flow-ex}. Initially, both ${\tt CPU0}$ and ${\tt CPU1}$ execute their memory read flows once simultaneously. Subsequently, ${\tt CPU0}$ executes its read flows two more times. The resulting trace is illustrated below 
\begin{equation}
\label{eq:trace}
\left(\{1, 3\}, 1, 2, 5, 1, 5, 6, 2, 4, 6, 2\right)
\end{equation}
where the numbers in the trace are the message indices as in Fig.~\ref{fig:flow-ex}(a). The concept of message sets allows us to represent the outcomes of these simultaneously executing flows.

Note that the ordering of the messages in the same set of a trace is unknown.  Given two messages $m_i$ and $m_j$ and a trace $\rho$, we define $m_i <_\rho m_j$, denoting that $m_i$ occurs before $m_j$ in $\rho$, if $m_i \in \varepsilon_i$, $m_j \in \varepsilon_j$, and $i<j$.  For simplification, a set of a single message $\{m\}$ is written as $m$. 

The objective of this method is to infer models that characterize the underlying message flows, similar to those shown in Fig.~\ref{fig:flow-ex}, from a given set of traces, $T$, to explain and understand the observed traces in $T$. Finite state automata are used to represent the extracted models.

\begin{definition}
A finite state automaton (FSA) is a tuple $\mathcal{M} = (Q, q_0, \Sigma, F, \Delta)$  where $Q$ is a finite set of states, $q_0 \in Q$ the initial state, $\Sigma$ a finite set of symbols, $F \subseteq Q$ the set of accepting states, and $\Delta : Q \times \Sigma \to Q$ the transition function.
\end{definition}

In our method, each symbol in $\Sigma$ denotes a unique message found in an input trace, and $F = \{q_0\}$.  
Given an FSA $\mathcal{M}$, a \emph{flow execution scenario} is defined as $\mathcal{X} = (\mathcal{M}, \{q^i_j~|~ 1 \leq i\leq n\})$ where $q^i_j$ is a current state of the $i^{th}$ instance of $\mathcal{M}$.
Suppose $\rho$ is an input trace and $\mathcal{M}$ is the model inferred from $\rho$.
The initial flow execution scenario is $\mathcal{X}_0 = (\mathcal{M}, \{q^i_0)~|~ 1 \leq i\leq n\})$.
Then, for every message $m$ in $\rho$ from the beginning, it is \emph{accepted} by $\mathcal{X}$ if we can find an $q^i_j$ in $\mathcal{X}$ such that $\Delta(q^i_{j}, m, q^i_{j+1})$ holds for some state $q^i_{j+1} \in Q$.
After accepting $m$, we get a new execution scenario $\mathcal{X}'$ where $q^i_j$ is replaced with $q^i_{j+1}$.
Accepting a message~$m$ is denoted as $\mathcal{X} \xrightarrow{m} \mathcal{X}'$. 
Naturally, accepting a message set~$\varepsilon$ is denoted as $\mathcal{X} \xrightarrow{\varepsilon} \mathcal{X}'$.   
The trace $\rho = (\varepsilon_0, \varepsilon_1, \ldots, \varepsilon_n)$ is \emph{accepted} by an execution scenario $\mathcal{X}$ if there is a sequence of its instances $\mathcal{X}_0, \mathcal{X}_1, \ldots, \mathcal{X}_{n+1}$ such that for each $0 \leq i \leq n$, $\mathcal{X}_i \xrightarrow{\varepsilon_i} \mathcal{X}_{i+1}$ where 
$\mathcal{X}_0$ is the initial execution scenario of $\mathcal{X}$.
The presence of multiple instances of the model $\mathcal{M}$ in an execution scenario indicates that the trace results from the concurrent execution of multiple message flows.
The goal of the proposed model synthesis method is \emph{to infer a model $\mathcal{M}$ from a set of traces such that an execution scenario $\mathcal{X}$ can be found to accept the trace $\rho$.}

\begin{figure}
    \centering
    \includegraphics[width=0.25\textwidth]{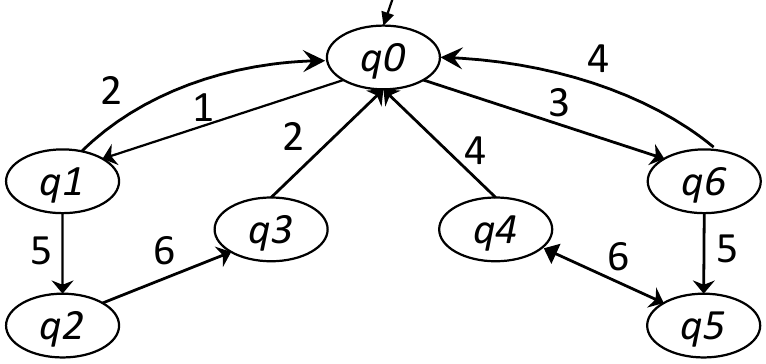}
    \caption{A FSA model for message flows in Fig.~\ref{fig:flow-ex}.}
    \label{fig:model-1}
\end{figure}

For the trace $\rho$ in $(\ref{eq:trace})$, an FSA model $\mathcal{M}$ that could be inferred is shown in Fig.~\ref{fig:model-1}. 
In this model, each path represents a particular branch of a message flow, \eg path $q_0,q_1, q_0$ for flow branch $(1,2)$. 
An execution scenario $\mathcal{X}$ that accepts $\rho$ in~(\ref{eq:trace}) is $(\mathcal{M}, \{q^i_0), 0 \leq i \leq 3\})$.

The message flows depicted in Fig.~\ref{fig:flow-ex} specify both causality relations among messages and their temporal relations. Temporal relations can be derived from causality relations; for instance, when message $\tt(CPU0:Cache:rd\_req)$ occurs, it triggers either $\tt(Cache:CPU0:rd\_resp)$ or $\tt(Cache:Mem:rd\_req)$ to occur later. To capture this, we introduce the concept of \emph{structural causality} based on the observation {that every message in a system execution trace is an output of a component in response to a previous input message.}

\begin{definition}
\label{def:causal}
Message $m_j$ is {causal} to  $m_i$, denoted as $\mathit{causal}(m_i, m_j)$, if $m_i{\tt .dest} = m_j{\tt .src}.$
\end{definition}

The causality defined above is termed \emph{structural causality}, distinguishing it from the functional causality described in the flow specifications. While functional causality implies structural causality, the reverse is not necessarily true. The objective of our method is to extract the functional or true causalities from the input traces.

\section{Limitation of the Existing Methods}
\label{sec:gap}

\begin{figure}
    \centering
    \includegraphics[width=3.4in]{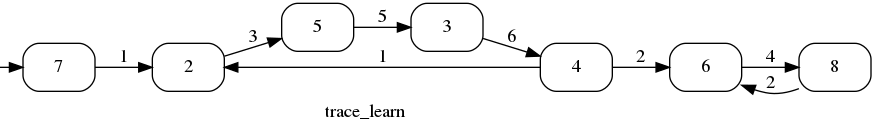}
    \caption{FSA model produced using the method in~\cite{natasa2020} for trace (\ref{eq:tr-2}).}
    \vspace{-5pt}
    \label{fig:trace2model-output}
\end{figure}

The difficulty in inferring models from communication traces lies in the concurrent execution of multiple message flows, making it challenging to correlate messages correctly for the respective flows based on temporal dependencies.
Note that $(1,5,6,2)$ and $(3,5,6,4)$ are two flows specified in Fig.~\ref{fig:flow-ex}. 
\begin{equation}
\label{eq:tr-2}
(1,3,5,6,1,3,5,6,2,4,2,4)
\end{equation}
Consider trace~(\ref{eq:tr-2}) obtained from executing the aforementioned message flows two times each in an interleaved manner. In this trace, messages $1$ and $3$ appear to exhibit strong temporal dependency, but they are actually unrelated according to the specification in Fig.~\ref{fig:flow-ex}. Existing methods do not successfully extract actual message flows because they rely on temporal dependencies to identify interesting patterns.
For instance, using $\mathit{Perracotta}$ in~\cite{Yang:2006}, we can extract a sequential pattern $(1, 3, 5, 6)$, which is highly confusing as it suggests that the memory read operations initiated by ${\tt CPU0}$ would always occur before the flow initiated by {\tt CPU1}. However, such a pattern does not exist in the message flows specified in Fig.~\ref{fig:flow-ex}. Fig.~\ref{fig:trace2model-output} shows the FSA model produced using the method in~\cite{natasa2020}. Although this model perfectly accepts the trace (\ref{eq:tr-2}), it suffers from the same problem.

In our method, the FSA model extracted from the traces must satisfy the structural causality relation for every two consecutive messages. Consequently, the sequence $(1, 3, 5, 6)$ would not be extracted as a model of the above trace since $\mathit{causal}(1, 3)$ does not hold. Applying the structural causality during the model extraction process is straightforward, as it solely relies on the basic structural information of a system design, which can be readily captured in messages.
\section{Model Synthesis}
\label{sec:method}

Algorithm~\ref{algo:flowminer} shows the outline of our method. It takes as an input a set of traces $T$ and produces an FSA model $\mathcal{M}$ such that an execution scenario of $\mathcal{X}$ can be found to accept $\rho$ as described in Section~\ref{sec:background}. The method involves three main steps: building and annotating the causality graph with information from $T$ (lines 3-5), generating the constraint problem $P$ from the causality graph (line 6), and extracting the model from the satisfying solutions for $P$ (lines 7-8). These steps are explained in the following sections. 

\setlength{\textfloatsep}{0pt}
\IncMargin{.5em}
\begin{algorithm}[tb]
\caption{\textbf{AutoModel}}
\label{algo:flowminer}
\textbf{Input:} A set of traces $T$\\
\textbf{Output:} {an FSA $\mathcal{M}$}\\
\SetAlgoNoLine
Extract unique messages in $T$ into $M$\;
Build the causality graph $G$ from $M$\;
Annotate $G$ \wrt $T$\;
Generate constraint satisfaction problem $P$ from $G$\;
Find a solution $sol$ of $P$ using a SMT solver\;
Derive an FSA $\mathcal{M}$ from $sol$\;
\end{algorithm}
\DecMargin{.5em}

\subsection{Building Causality Graph}
\label{sec:cg}

Given an input trace $\rho$, we scan it to collect all unique messages. A message is considered unique if at least one of its three attributes differs from all other messages already collected. We can identify initial and terminal messages for each message flow during the message collection process. Specifically, a message $m$ is classified as an initial message if there is no other message $m'$ in $\rho$ such that
\begin{equation}
\label{eq:start-m}
m' <_\rho m \mbox{ and } \mathit{causal}(m', m) 
\end{equation}
Message $m$ is a terminal if is no message $m'$ in $\rho$ such that
\begin{equation}
\label{eq:end-m}
    m <_\rho m' \mbox{ and }   \mathit{causal}(m, m') 
\end{equation}
By scanning the trace from the beginning, we can find all initial messages by checking condition~(\ref{eq:start-m}). Similarly, all terminal messages can be found by scanning the trace from the end and checking condition~(\ref{eq:end-m}). 

We create a causality graph $G$ from the collected messages to capture all possible structural causalities among them. The graph is directed and acyclic, with root nodes labeled as initial messages from the set $Start$ and terminal nodes labeled as terminal messages from the set $End$. The remaining nodes are labeled with messages that are neither initial nor terminal. Edges in the graph represent structural causality relations between messages, as defined in Definition~\ref{def:causal}.

\begin{figure}
    \centering
    \begin{tabular}{cc}
        \begin{minipage}{1.6in}
        \[
        \begin{array}{lcl}
         M & = & \{1,2, 3,4,5,6\}\\
         Start & =& \{1, 3\}\\
         End &=& \{2,4\}
        \end{array}
        \]
        \end{minipage}
         &
         \begin{minipage}{1.6in}
             \includegraphics[width=1.6in]{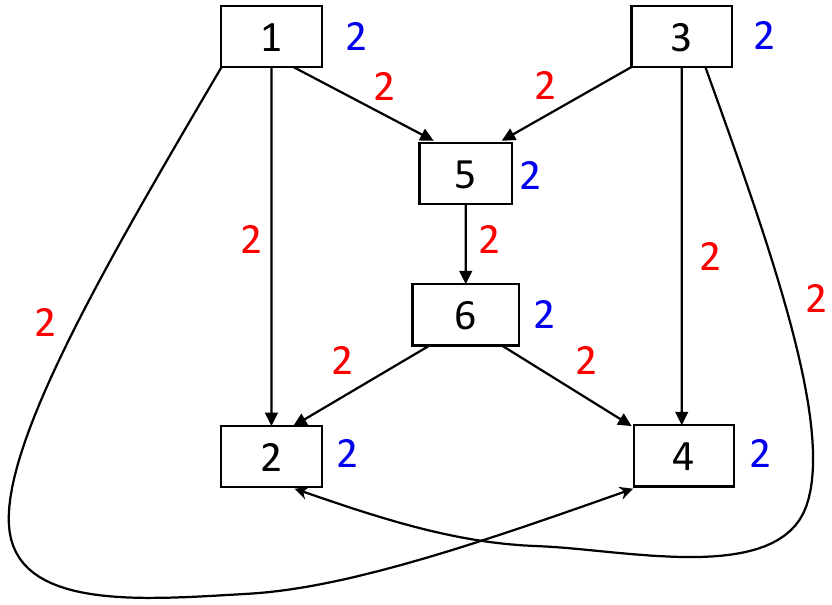}
        \end{minipage}
        \\
         (a) & (b)
    \end{tabular}
    \caption{Messages and causality graph constructed from trace $(\ref{eq:tr-4})$.}
    \label{fig:cg-ex}
\end{figure}

After constructing the causality graph, we scan the trace again to determine the support for nodes and edges. The support of a node is the number of instances of the labeled message in the trace. For each edge with head node $h$ and tail node $t$, its support is calculated as the number of instances of $m_t$ associated with an instance of $m_h$ such that $m_h <_\rho m_t$ and $m_h$ is not matched with any other instances of $m_t$ in the trace. $m_h$ and $m_t$ are messages labeled for nodes $h$ and $t$, respectively.
Consider the example trace below 
\begin{equation}
\label{eq:tr-4}
(1,3,5,6,4,2,3,1,5,6,2,4).
\end{equation}  
Fig.~\ref{fig:cg-ex} illustrates the collected messages and the corresponding causality graph for the given trace. The nodes in the causality graph are labeled with indices of messages, as defined in Fig.~\ref{fig:flow-ex}(a). Node supports are indicated by numbers in blue, while edge supports are shown in red. In this specific example, both node and edge supports are equal to $2$. Calculating node supports is straightforward while determining edge supports is a bit more involved.
Let $x\to y$ denote an edge from node $x$ to $y$. Take the edge $1 \to 5$ as an example. To find its support, we first find all instances of message~$5$, which are $2$. Then, for each instance of message~$5$, we check if there is an unmatched instance of message~$1$ before message~$5$ in the trace. This condition is met in the above trace for both instances of message~$5$, therefore, the support for edge $1 \to 5$ is $2$ for this particular example.

\subsection{Generating and Solving Consistency Constraints}
\label{sec:constr-solve}
The constructed causality graph contains potential models for the input trace. However, to derive a valid model, we must ensure that the causality graph is free from inconsistencies. An inconsistency can arise when the support of a node does not match the support of its incoming or outgoing edges. For instance, consider node5 and its incoming edges in Fig.\ref{fig:cg-ex}. The node support is $2$, while the total support of its incoming edges is $4$. This inconsistency occurs due to the ambiguities when finding the edge supports. In this example, we encounter uncertainties in determining how message5 should be correlated with other messages, leading to multiple possibilities. Consequently, we calculate supports for all pairs of messages ending with message5, \textit{e.g.}, $(1, 5)$ and $(3,5)$, that satisfy the causality requirement. As a result, message~5 is counted more than it should be.
   
The goal of our method is to identify a set of edges whose supports align with the node supports. To accomplish this, we formulate and solve a constraint satisfaction problem generated from the causality graph. Let $sup(\cdot)$ denote the support of a node or an edge, and $c(n\to n')$ represent a variable representing the \emph{consistent} support of the edge $n\to n'$. From the causality graph, we establish the following constraints.

\begin{enumerate}
    \item For each node $n$, and all its outgoing edges $n\to n'$, create a constraint
        \[sup(n) = \sum_{\mbox{all } n\to n'} c(n\to n') \]
    \item For each node $n'$ and all its incoming edges $n \to n'$, create a constraint
            \[sup(n') = \sum_{\mbox{all } n\to n'} c(n\to n') \]
    \item For each edge $n\to n'$, create a constraint
            \[0 \leq c(n\to n') \leq sup(n\to n') \]
\end{enumerate}
The first two constraints ensure consistency between each node and its incoming and outgoing edges. The third constraint accounts for the uncertainty surrounding the exact edge support when scanning the trace for instances of an edge, limiting it to be no greater than the supports directly obtained from the trace. These constraints collectively form the constraint satisfaction problem $P$, from which models can be derived based on its satisfying solutions. The objective of solving $P$ is to identify a set of consistent edge supports $c(n\to n')$ that simultaneously satisfy all constraints in $P$.

\subsection{Deriving Model}  
\label{sec:model-gen}

\begin{figure}
    \centering
    \begin{tabular}{cc}
    \begin{minipage}{1.5in}
    \includegraphics[width=1.2in]{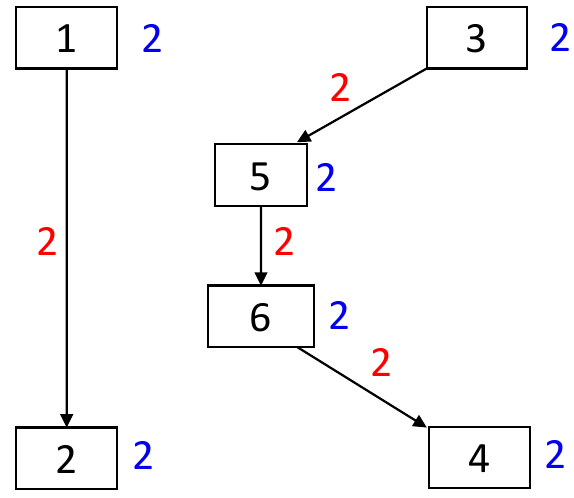}
    \end{minipage}
    &     
    \begin{minipage}{1.5in}
    \includegraphics[width=1.2in]{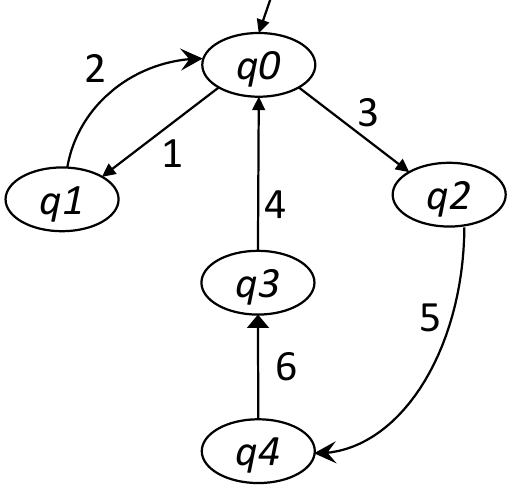}
     \end{minipage}
\\
      (a)   & (b) 
    \end{tabular}
    \caption{(a) The modified causality graph showing a consistent solution derived from the trace $(\ref{eq:tr-4})$, and (b) the corresponding FSA model.}
    \label{fig:cg-sol-1}
\end{figure}

Once the constraint problem $P$ is formulated, it is inputted into a constraint solver to find a solution. A solution $\mathit{sol}$ of $P$ is a set of edges in the causality graph $G$ for which their supports are consistent with the node supports,
\[
\{n\to m~|~n\to m \mbox{ in } G, \mbox{ and } \mathit{c}(n\to m) > 0 \mbox{ in } \mathit{sol}\}.
\]   
The obtained solution can be visualized as a modified causality graph with the edge supports as returned from the solver. An example solution for the causality graph shown in Fig.~\ref{fig:cg-ex} is illustrated in Fig.~\ref{fig:cg-sol-1}(a), and the corresponding FSA model is presented in Fig.~\ref{fig:cg-sol-1}(b).
Note that in Fig.~\ref{fig:cg-sol-1}(a), edges from Fig.~\ref{fig:cg-ex} with zero-supports in the solution are removed. 

For the constraint problem $P$ generated in the previous step, a substantial number of models can be derived. Deriving a minimal model is a common practice in model synthesis methods~\cite{Lang:1998:EDSM, Ulyantsev:2011, Heule:2013, natasa2020}. However, finding the minimal model is an NP-hard problem. Thus, in this step, our objective is to efficiently generate a reduced solution, which may not necessarily be minimal but remains more understandable to humans and easier to use in activities such as verification and testing.
In our method, we query the solver to return a set of solutions $S= \{sol~|~sol \models P\}$. Then, for each solution $\mathit{sol} \in S$, and for each edge $n\to m$ with non-zero support in $\mathit{sol}$, a new constraint is generated where its support is set to $0$.
After incorporating this constraint into the solver, if the solver becomes infeasible, $\mathit{sol}$ is returned as a candidate. Otherwise, the above step is repeated for the reduced solution $\mathit{sol}'$. Ultimately, from the set of reduced solutions, we select the one with the smallest number of edges with non-zero support and return it to the user. The model extraction method is illustrated in Algorithm~\ref{algo:model-extract}.
The model in Fig.~\ref{fig:cg-sol-1} shows an example model extracted using Algorithm~\ref{algo:model-extract}. Note that sequence $(1, 3)$ or $(2, 4)$ is not included in the derived model, compared with the results mined using previous methods as described in Section~\ref{sec:gap}. 

After obtaining a reduced solution as described above, an FSA can be constructed from it. This step is straightforward and we skip further explanation.

\subsection{Model Synthesis from Multiple Traces}
\label{subsec:multiple}

\begin{figure}
    \centering
    \includegraphics[width=3.55in]{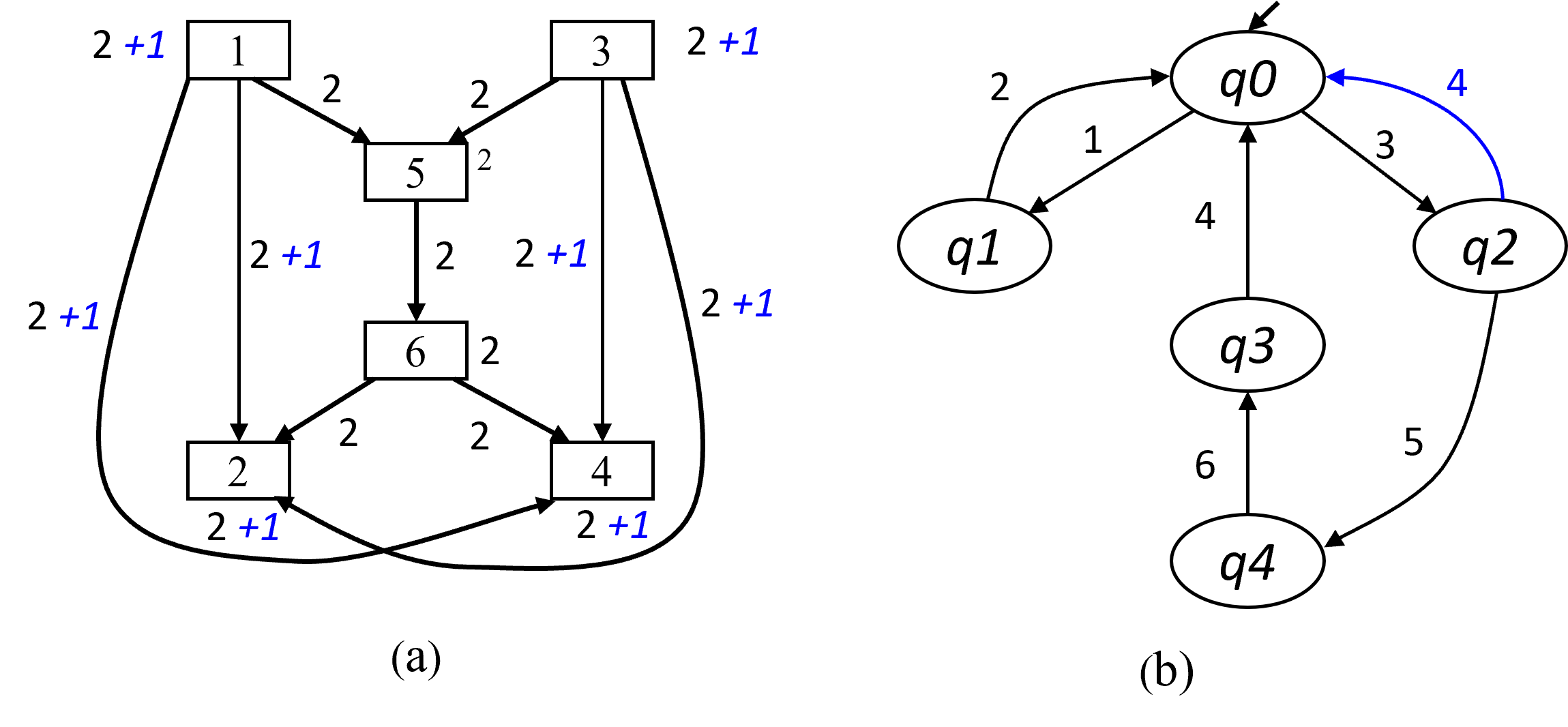}
    \caption{(a) Causality graph constructed from traces in~$(\ref{eq:tr-4})$ and $(\ref{eq:tr-5})$, and (b) a FSA model generated from the causality graph in~(a) where a new sequential pattern $(3, 4)$ is discovered.}
    \label{fig:muti_trace_cas}
\end{figure}

The above description of the model synthesis considers a single trace. 
However, our method can be readily extended to process a set of traces as follows. 
Initially, the support of each node and each edge of the causality graph are set to zero. 
For each trace $\rho \in T$, the supports for each node and each edge in the causality graph are obtained, and then added to the existing supports of corresponding nodes and edges, respectively. 
The final node supports are the sum of node supports obtained from individual traces. 
The final edge supports are computed similarly.
Once all traces are processed, an inferred model from this causality graph is consistent with all input traces.

Consider trace $(\ref{eq:tr-4})$ in the previous section and $(\ref{eq:tr-5})$ below for a set of traces. 
\begin{equation}
\label{eq:tr-5}
(1, 3, 2, 4)
\end{equation}
Fig.~\ref{fig:muti_trace_cas}(a) shows the causality constructed for this trace set.
Compared to the causality graph shown in Fig.~\ref{fig:cg-ex}(b), the support of every node of this causality graph is the total number of instances of the corresponding labeled messages in both traces.    
Similarly, four edges have incremented supports in Fig.~\ref{fig:muti_trace_cas}(a).
A new model is extracted from this causality graph and is shown in Fig.~\ref{fig:muti_trace_cas}(b).
This new model is similar to Fig.~\ref{fig:cg-sol-1}(b), but with a new edge $(q_2, 4, q_0)$ included to reflect that a new message sequence pattern $(3, 4)$ is found.

\IncMargin{1.5em}
\begin{algorithm}[tb]
  \SetKwData{Left}{left}
  \SetKwData{Up}{up}
  \SetKwInOut{Input}{input}
  \SetKwInOut{Output}{output}
\caption{\textbf{ModelExtract}}
\label{algo:model-extract}

\Indm\Indmm
  \Input{Constraints $P$}
  \Input{Size limit of solutions $sz$}
  \Output{A reduced solution $\mathit{sol}$}
\Indp\Indpp
  \BlankLine
  $C = \emptyset$\;
  Find $S = \{\mathit{sol}~|~\mathit{sol} \models P\}$ of size $sz$\;
  \ForEach{$sol \in S$}{
    $sol' := \mbox{ReduceModel}(P, sol)$\;
    $C := C \cup \{ sol'\}$\;
  }
  Let $sol \in C$ with the minimal size\;
  return $sol$\;
\end{algorithm}
\DecMargin{1.5em}

\IncMargin{1.5em}
\begin{algorithm}[tb]
  \SetKwData{Left}{left}
  \SetKwData{Up}{up}
  \SetKwInOut{Input}{input}
  \SetKwInOut{Output}{output}
\caption{\textbf{ReduceModel}}
\label{algo:reduce-model}

\Indm\Indmm
  \Input{Constraints $P$}
  \Input{A solution $\mathit{sol}$}
  \Output{A reduced solution $\mathit{sol}'$}
\Indp\Indpp
  \BlankLine
    Get an edge $(n\to m)$ in $\mathit{sol}$\; 
    $P' := P \cup c(n\to m) =0$\;
    \If{$P'$ is unjustifiable}{\Return $\mathit{sol}$}
    Get a new solution $\mathit{sol}' \models P'$\;
    \Return ReduceModel($P'$, $\mathit{sol}'$)\;
\end{algorithm}
\DecMargin{1.5em}

\begin{figure}
    \centering
    \begin{tabular}{cc}
    \begin{minipage}{1.3in}
    \centering\includegraphics[width=1.2in]{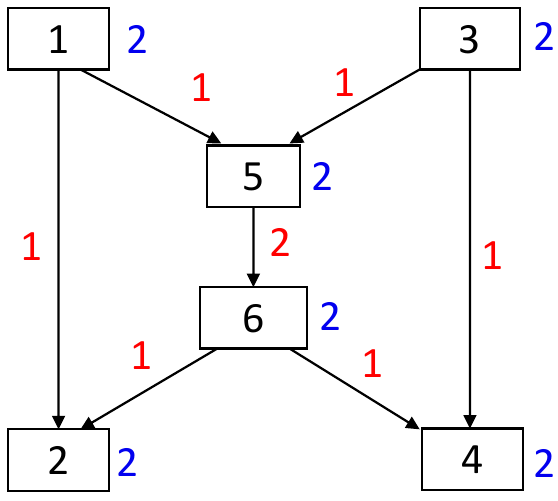}
    \end{minipage}
    &     
    \begin{minipage}{1.5in}
    \centering\includegraphics[width=1.4in]{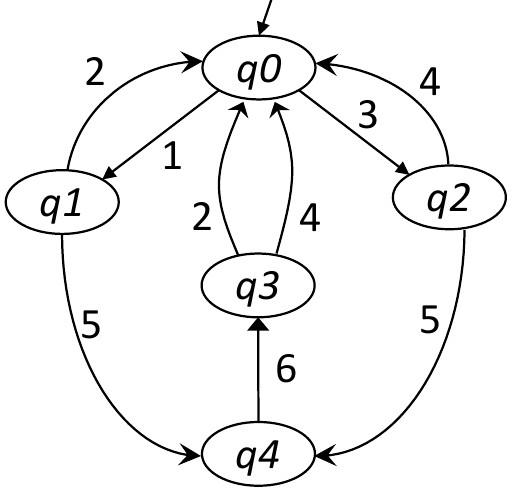}
     \end{minipage}
     \vspace*{6pt}
\\
      (a)   & (b) 
    \end{tabular}
    \caption{(a) The modified causality graph showing a consistent solution derived from  trace $(\ref{eq:tr-4})$ considering a window constraint, and (b) the corresponding FSA model derived from the new solution in (a).}
    \label{fig:cg-sol-2}
\end{figure}

\section{Optimization Techniques}
\label{sec:opt}

A critical task of our method is to extract true causality relations among messages from the input traces.  
On the other hand, the traces resulting from executing multiple message flows concurrently are highly unstructured; therefore, correctly correlating messages for mining is difficult.
As discussed, using temporal dependencies to approximate causality relations of messages often leads to invalid models being mined.
This section describes several techniques to accurately correlate messages for enhanced causality relations or extract certain true causality relations directly from the input traces.  

\subsection{Sliding Window Constraints}
We apply the traditional idea of the sliding window technique. Generally, the optimal size of the window is not known. For a given window constraint, the causality relation among the messages within the window is considered for causality graph building. A window slides from the first to last events on any trace or sub-trace. Algorithms~\ref{algo:reduce-model} returns a reduced FSA model consistent with the input trace. We adopt the sliding window technique aiming to filter out certain models that are likely unrealistic with respect to the input trace.

In the causality graph, the nodes represent messages exchanged in communications, and the edges represent the causal relations between two messages.
For such an edge $m_1 \to m_2$, its support is the count of occurrences of the sequence $(m_1, m_2)$ in a trace.
In the above algorithm, the edge support is computed without considering the distance between $m_1$ and $m_2$ in a trace. However, in practice, the distance between two messages can be used to evaluate how tightly they correlate by causality. The distance between two messages in a trace is measured by the number of messages that separate them. A higher distance indicates that the two messages are further apart in time from each other, suggesting that, in practice, they are less likely to be correlated by causality.

Considering the trace $(\ref{eq:tr-4})$ used to generate a consistent solution and model as shown in Fig.~\ref{fig:cg-sol-1}, we can impose a constraint to only correlate two messages and count them for edge support if their distance is no more than \emph{two}. This distance means no more than two other messages separate them in the trace. With this constraint, the supports for edges $(1,2)$ and $(3,4)$ become 1 instead of 2 in Fig.~\ref{fig:cg-ex}.
Additionally, the supports for edges $(1,4)$ and $(3,2)$ become 0, implying that sequences cannot exist in a solution. Therefore, the solution found in Fig.~\ref{fig:cg-sol-1} would be infeasible. 
However, when imposing the constraint, a new solution and model for the same trace are found and shown in Fig.~\ref{fig:cg-sol-2}. Despite the new model being larger than the one shown in Fig.~\ref{fig:cg-sol-1}(b), it is more meaningful in representing the behavior of the trace. The distance constraint excludes certain unrealistic causalities among messages, making the new model better suited for capturing the system's actual behavior.

Specifically, given a trace and two messages $m_{1,i}$ and $m_{2, j}$ where $i$ and $j$ are position indices of $m_1$ and $m_2$ in the trace, and $i < j$. 
Messages $m_{1, i}$ and $m_{2,j}$ are considered as correlating to each other by causality under a distance constraint $w$ if 
\begin{equation}
\label{eq:edge_sup_new}
j \leq i + w + 1~~\mbox{and}~~\mathit{causal}(m_{1,i}, m_{2,j}) \mbox{ holds.}
\end{equation}
The edge support for $(m_1, m_2)$ in the causality graph is the number of pairs of $m_1$ and $m_2$ that satisfy $(\ref{eq:edge_sup_new})$.   

\subsubsection{How to Select Distance Constraint} Considering a trace resulting from the concurrent execution of multiple systems flows, it is important to note that two causal messages in a flow may not occur one immediately after the other. 
For example, messages $1$ and $2$ in Fig.~\ref{fig:flow-ex} in trace $(1,3,5,6,4,2,3,1,5,6,2,4)$.
Therefore, selecting the distance constraint appropriately is a challenging task since a constraint of $0$ would miss true causality relations, while a large distance constraint would lead to irrelevant causality extractions, making the extracted model less meaningful. Therefore, choosing an appropriate distance constraint is an important but difficult problem as it involves many factors such as system structure, how flows are executed,\etc. In practice, user insights about the target design would be invaluable for choosing an appropriate distance. 
We propose an iterative heuristic to automatically find a model with minimal distance without requiring user input. The process is as follows
\begin{enumerate}
    \item Initially, the distance constraint is set to $0$.
    \item If the model synthesis algorithms can construct a model, that model is returned.
    \item Otherwise, increment the distance constraint by $1$, and repeat the previous step.
\end{enumerate}
The above process iterates until a model is successfully inferred. This heuristic allows the method to automatically determine an appropriate distance constraint for the given input traces.

\subsection{Trace Slicing on Message Attributes}
Modern complex SoCs that can run operating systems trivially consist of multiple cores and can exert parallelism during their executions. Therefore, multiple instances of different flows or the same flow can be active simultaneously. Due to this high degree of concurrency, accurately calculating edge support for the causality graph is a major challenge.  
Work in~\cite{yuting_lstm} shows that causality slicing on micro-architectural information can improve message correlation. 
In this work, we explore other possible message attributes to find better message correlations, thus, more accurate edge support calculations for causality graphs. 
For example, if two messages are not causal to each other, we do not increase the support of the respective edge.
Similarly, if two events have different message attributes, such as memory address, they are unlikely to be part of the same flow instance; thus, we can skip counting the support for them. In this section, we use gem5 to describe how micro-architecture attributes can be used to slice the traces so that messages are correlated in the sliced traces more accurately. 

\paragraph{Packet Attributes} Packets are the base units of transactions in the gem5 simulation. A packet encapsulates a transfer between two objects or sim objects in the memory system (e.g., the L1 and L2 cache). 
When a master object (CPU) makes a request, it gets conveyed to the ultimate destination by several packets. 
Therefore, tracking packets through different communication links provides a reliable request-tracking method. 
Typical packet attributes in gem5 include {\tt cmdIndex, masterId, ContextID, Request::depth, packetId, packet.addr, packet.size, packet.flags, packet.tick}, \etc. 
More details about these attributes can be found in the gem5 official packet class reference~\cite{pkt_class}. 
Later in Section~\ref{sec:results}, we show that the accuracy can be improved for the models inferred from traces sliced with respect to the following packet attributes. 

\paragraph{Memory Address} A message can possess a memory address, which serves two purposes: routing the packet to its intended destination when the destination is not explicitly specified and facilitating the processing of the packet at the target. This address is typically derived from the physical address of the requesting object. However, in certain scenarios, such as accessing a fully virtual cache before address translation occurs, it may be derived from the virtual address instead. It's important to note that the address in the message may not be identical to the address of the original requesting object. For instance, in the case of a cache miss, the packet address may refer to the memory block that needs to be fetched rather than the original requesting object's address. Therefore, an address mapping is carried out on the address-sliced traces to correct the slicing, such as the CPU generates an address $p$, then cache generated block address $q$, will be same as $q = \lfloor p/N \rfloor$, whereas $N$ is the size of the cache line.

\paragraph{Packet Id} gem5 messages have different kinds of attributes. {\tt packet id} is a public {\tt const } variable in of packet class. It is unique for each requesting object in the simulation.  {\tt Packet id} is assigned by the requesting component of the simulating model. Packet id is unique at any execution time, and all the simultaneously active flow instances have different {\tt packet id}s, one for each instance. Therefore, some {\tt packet id} could be shared by different flows at different times but not simultaneously. {\tt packetID} is important for packet tracking as it uniquely identifies each packet throughout its journey in the simulated system. It allows gem5 to accurately monitor and analyze the movement, transformations, and interactions of packets, facilitating performance analysis and debugging of the simulated system.

\paragraph{context Id} The {\tt contextID} is a field present in the packets used for distinguishing different execution contexts or threads within a simulated system. It is particularly relevant in multi-threaded simulations where multiple threads or processes are running concurrently. The {\tt contextID} serves as a unique identifier associated with each thread or process. It helps gem5 track and differentiates the packets generated by different execution contexts within the simulated system. By including the {\tt contextID} in the packets, gem5 can maintain the proper ordering and handling of requests and responses from different threads or processes.

In this work, we demonstrate the effectiveness of trace slicing using the above message attributes available in gem5. {\tt packetID} and {\tt contextID} can play essential role for slicing gem5 packet traces. As unique identifiers, packet ID, and context ID enable accurate analysis of packet movement and interactions. Distinguishing messages based on packet ID prevents false counting of edge support in the causality graph. Similarly, globally unique context IDs prevent erroneous edge support calculations, ensuring more accurate trace analysis in gem5. In practice, the same idea can be readily applied where the designers can identify design-specific attributes to assist trace slicing.

\section{Evaluation Method}
\label{sec:eval}




Previous work on sequential pattern mining~\cite{Le:2018:DSM:3213846.3213876, workflow_int_traces, quark} often uses precision/recall/F1-measure to evaluate the quality of the mined results. 
However, they rely on the availability of ground truth specifications to compute those scores.
Deriving minimal models is widely adopted in model synthesis methods~\cite{Lang:1998:EDSM, Ulyantsev:2011, Heule:2013}. Additionally, a minimized model is favored due to its enhanced human comprehensibility and ease of use in various activities such as verification and testing.
However, the sizes of the inferred models do not provide a reliable indicator of how well the inferred models can be used to understand and explain the input traces. 
This paper proposes a new evaluation approach, \emph{acceptance ratio} (AR), to overcome those issues.
Given a model $M$ inferred from a trace $\rho$, an acceptance ratio is a fractional number indicating the number of messages in $\rho$ accepted by $M$ versus the total length of $\rho$.
A good model should have a high acceptance ratio over all traces where the model is inferred.

Algorithm~\ref{algo:evaluationAlgorithm} shows how the acceptance ratio for an FSA model and a trace as inputs is computed. 
The idea behind this algorithm is based on the trace acceptance described in Section~\ref{sec:background}.
If the message under examination is one of the initial messages, a new instance of the sub-FSA corresponding with the same initial message is added to the active FSAs array. 
On the other hand, if the message under examination is accepted by one of the instances in the active FSAs array, the corresponding instance will be updated with the new event. 
In both of these cases, the message will be added to the accepted messages array and, if the message cannot be accepted by either of these cases, it means that the message could not be accepted by the inferred FSA.
After all of the messages in the trace are examined, the acceptance ratio is calculated based on the number of accepted messages to the total number of messages in the trace.
For example, using the inferred FSA in Fig.~\ref{fig:model-1} and trace~(\ref{eq:trace}) as the inputs, the final acceptance ratio will be equal to 1 as all of the messages could be accepted by the inferred FSA. 
However, using the inferred FSA in Fig.~\ref{fig:cg-sol-1}(b) and trace~(\ref{eq:trace}), the final acceptance ratio will be 0.83, as two messages (5 and 6) could not be accepted by the input FSA. 
As can be seen in this example, larger FSAs are capable of modeling the system behavior better than the smaller FSAs. 


\IncMargin{1.5em}
\begin{algorithm}[tb]
  \SetKwData{Left}{left}
  \SetKwData{Up}{up}
  \SetKwInOut{Input}{input}
  \SetKwInOut{Output}{output}
\caption{\textbf{Evaluation Algorithm}}
\label{algo:evaluationAlgorithm}

\Indm\Indmm
  \Input{Trace $\rho$}
  \Input{Inferred FSA $\mathcal{M} = \{Q, q_0, \Sigma, F, \Delta\}$}
  \Output{Acceptance Ratio $AR$}
\Indp\Indpp
    $Accept = \emptyset$\;
    $X = \emptyset$\;
  \ForEach{$i \in [0, |\rho|]$}{
    $m \gets \rho[i]$\;
    \If{$m$ is a start message} {
        find $q_1$ s.t. $\Delta(q_0, m, q_1)$ holds\;
        Create and add a model instance $(\mathcal{M}, q_1)$ to $X$\;
        Add $m$ to $Accept$\;
    }
    \ElseIf{$\exists (\mathcal{M}, q) \in X,~\Delta(q, m, q')$ holds}{
        $X \gets X \cup \{(\mathcal{M}, q')\} - (\mathcal{M}, q)$\;
        Add $m$ \mbox{to} $Accept$\;
    }
  }
  \Return $AR = |Accept|/|\rho|$\;
\end{algorithm}
\DecMargin{1.5em}



\begin{figure}
    \centering
    \includegraphics[width=.25\textwidth]{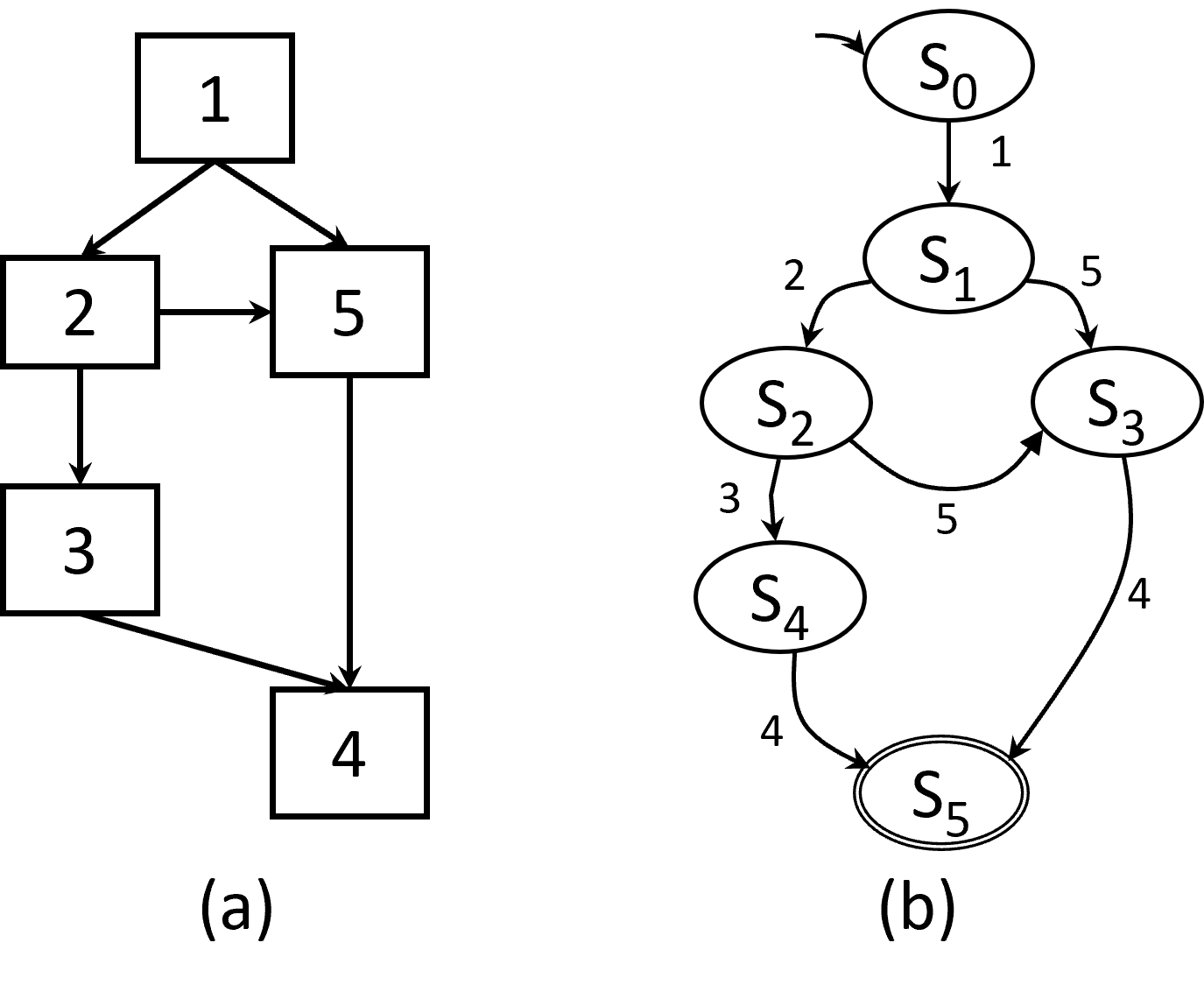}
    \caption{A consistent causality graph (a) and corresponding FSA (b). }
    \label{fig:challengeExample}
\end{figure}

\medskip
\hspace{-15pt} {\bf Challenges in FSA evaluation:}
\label{subsub:fsa_challenges}
Using FSAs to validate the traces comes with a challenge, and it is caused by the non-determinism in the choices of FSA instances. This means that when a new event is being checked with the FSA, there may be more than one possible active instance of the inferred FSA that can accept the event. An example of this challenge could be found in the inferred FSA in Fig.~\ref{fig:challengeExample}(b) and (1, 1, 2, 1, 5, 5, 4, 4, 3, 4) as the input trace file. There are two possible outcomes for this example, \{(1, 2, 3, 4), (1, 5, 4), (1, 5, 4)\} and \{(1, 2, 5, 4), (1, 5, 4), (1)\} which are respectively acceptable and unacceptable (incomplete). Thus, choosing an FSA instance to assign the new event, can determine if the trace will be accepted or not, and choosing the best FSA instance is the main challenge.

\section{Experimental Results}
\label{sec:results}

We implement the method described in this paper in Python and use the Z3 SMT solver~\cite{z3prover} to solve the constraint problems. We refer to this implementation as {\sf AutoModel} and evaluate it in two experiments.
All the experiments were conducted on an Intel Core i7-12700 CPU operating at a clock speed of 4.8 GHz with 32 GB of main memory. The source code is available at~\cite{project_repo}.

\subsection{Preparing SoC Traces}
The model synthesis method is developed to extract models from real execution traces. Complex SoCs with multiple compute units, peripherals, and multilevel memory systems with advanced interconnects such as Ruby, Garnet2.0, and memory controllers that can run complex workloads such as operating systems are the target use cases for this endeavor.  However, accessing such complex execution traces of real operating scenarios is often not feasible due to its proprietary nature. Therefore, we evaluated our method on two different sets of traces. The first set of traces are synthetic traces. The main benefit of synthetic traces is that the underlying ground truth message flows are available. Therefore, we can compare the models generated by our method against the ground truth flows to better evaluate the qualities of the generated models.
To evaluate our method more practically, we use the widely known architectural platform gem5~\cite{gem5} to build more realistic SoC models. We revise the gem5 internals to capture the messages on the communication links of our interest when the models are simulated. More details of the generation of the above traces are described in the sections below.

\subsubsection{Synthetic Trace Generation}
In the first experiment, we collect 10 message flows similar to the flows in Fig.~\ref{fig:flow-ex} from our industry partners. These flows are abstracted from system-level communication protocols used in real industry designs, therefore much more sophisticated.
Each flow consists of a number of branches, resulting in a total number of $64$ message sequences to specify various system communication scenarios such as cache-coherent memory accesses,  upstream read/write, power management, etc.  Although greatly simplified, these message flows capture essential communicating behaviors among various components in a typical system design, including CPUs, caches, interconnect, memory controllers, peripheral devices, etc. 

\begin{table}[!ht]
\centering
\caption{Synthetic traces features }
\renewcommand{\arraystretch}{1.3}
\begin{tabular}{|l|c|c|cc|}
\hline
\multicolumn{1}{|c|}{\multirow{2}{*}{Trace}} & \multirow{2}{*}{Length} & \multirow{2}{*}{\#Msg} & \multicolumn{2}{c|}{Non-zero edges} \\ \cline{4-5} 
\multicolumn{1}{|c|}{} &      &                     & \multicolumn{1}{c|}{Unsliced} & Window Sliced \\ \hline
{\sf Small-5}                & 920  & \multirow{3}{*}{34} & \multicolumn{1}{c|}{214}      & 210           \\ \cline{1-2} \cline{4-5} 
{\sf Small-10}               & 1840 &                     & \multicolumn{1}{c|}{214}      & 214           \\ \cline{1-2} \cline{4-5} 
{\sf Small-20}               & 2754 &                     & \multicolumn{1}{c|}{214}      & 214           \\ \hline
{\sf Large-5}                & 1666 & \multirow{3}{*}{60} & \multicolumn{1}{c|}{665}      & 508           \\ \cline{1-2} \cline{4-5} 
{\sf Large-10}               & 3152 &                     & \multicolumn{1}{c|}{667}      & 569           \\ \cline{1-2} \cline{4-5} 
{\sf Large-20}               & 7656 &                     & \multicolumn{1}{c|}{668}      & 635           \\ \hline
{\sf Multiple}               & 5514 &          34           & \multicolumn{1}{c|}{159}      & 159           \\ \hline
\end{tabular}
 \label{tab:synt_analysis}
\end{table}


We developed a transaction-level simulation model where components communicate according to the specified protocols in the message flows. In this simulation, each master component initiates a message flow instance randomly by generating a message and sending it to the next component. The simulation considers only the command and destination address in the messages, omitting message payloads. To reflect the concurrent nature of modern system designs, we modeled the interconnect component as a switch network, enabling concurrent processing of multiple messages from different sources. All components run concurrently without global synchronization in the simulation.

Two sets of traces are collected by simulating a transaction-level model using two different configurations. The first set of traces, referred to as "small traces," is generated when only CPUs are allowed to initiate cache coherent downstream read flows, while all other components are configured to only react to incoming messages. The second set of traces, referred to as "large traces," is generated with all components in the model enabled to initiate message flows. For each configuration, the system model is simulated three times, resulting in three traces of each type. During the simulation runs, a fixed number of flow instances are activated.


The flow instance numbers used in simulation runs are set to $5$, $10$, and $20$ as indicated in the first column of Table~\ref{tab:synt_analysis}.  
A larger flow instance number used in simulation leads to a longer trace: the second row and the number of unique messages in each trace are shown in the third row. In addition to those traces, a third set of traces is prepared, combining all the {\sf Small} traces. This third set, referred to as {\sf Multiple} in the following paragraph, is prepared to apply the model synthesis from multiple traces. Synthetic traces help test the accuracy of the synthesized FSA. This allows us to differentiate between an overly accepting FSA and one accurately representing valid trace execution behavior.

\subsubsection{gem5 Trace Generation}
\begin{figure}
    \centering
    \includegraphics[width=3.5in]{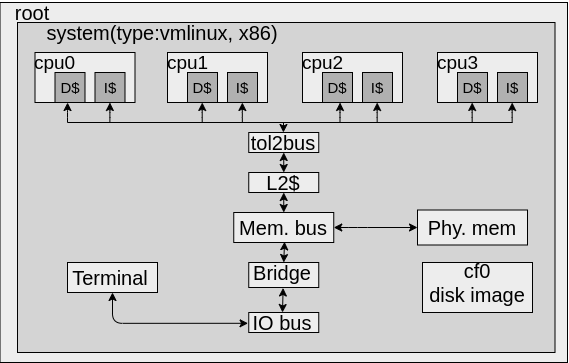}
    \caption{gem5 Simulation Setup.}
    \label{fig:gem5fs}
\end{figure}

 gem5 is a widely-used open-source computer architecture simulator that can run in one of two modes: \textit{Full System simulation (FS)} and \textit{Syscall Emulation (SE)}. 
 Fig.~\ref{fig:gem5fs} shows the high-level diagram of a multicore SoC model.
 It contains four x86 TimingSimpleCPU cores, each with a private data cache (64kB) and a private instruction cache (16kB).  
 All cores share a 256kB level 2 cache. 
 There is also a DDR4\_2400\_16x4 memory controller with an address space of up to 4GB. These IPs are connected through high-speed concurrent interconnects that can handle multiple requests simultaneously. 
 The IP blocks in this SoC communicate with each other to realize various operations. We execute different binaries as workloads on this bare-bone gem5 setup to extract realistic and sophisticated traces.
 We instrument the gem5 model with 20 communication monitors to observe the {\tt packet}s (unit of communication) over different communication links. 
 Table~\ref{tab:gem5_traces} describes the traces from simulating the model in both modes.



\begin{table}[!t]
\renewcommand{\arraystretch}{1.3}
\caption{gem5 Traces~\cite{rubel-isvlsi} features. The length of the traces is measured in millions.}
\centering
\begin{tabular}{|ll|c|c|c|}
\hline
\multicolumn{2}{|l|}{\textbf{Trace}}                                      & \multicolumn{1}{l|}{\sf Kernel} & \multicolumn{1}{l|}{\sf Threads} & \multicolumn{1}{l|}{\sf Snoop} \\ \hline
\multicolumn{2}{|l|}{\textbf{Length}}  & 8596.5 & 7.7 & 0.55 \\ \hline
\multicolumn{2}{|l|}{\textbf{\#Msg}}   & 147     & 154  & 115   \\ \hline
\multicolumn{1}{|l|}{\multirow{4}{*}{\textbf{\begin{tabular}[c]{@{}l@{}}Non-zero\\ edges\end{tabular}}}} & Unsliced & 1986                    & 2412                         & 1541                       \\ \cline{2-5} 
\multicolumn{1}{|l|}{} & Window sliced & 653     & 821  & 572   \\ \cline{2-5} 
\multicolumn{1}{|l|}{} & Architectural & 279     & 345  & 250   \\ \cline{2-5} 
\multicolumn{1}{|l|}{} & Total sliced & 271     & 340  & 239   \\ \hline
\end{tabular}
\label{tab:gem5_traces}
\end{table}

The full system simulation trace {\sf Kernel} captures critical behavior while booting a real operating system. It is generated when booting the Linux kernel
4.1.3. However, no other workload is executed except the kernel. The {\sf Snoop} trace captures the cache coherent interconnect behavior for snooping activity. It activates a simple snoop protocol implemented
for the classical multi-level memory system. 
The {\sf Threads} trace is generated when two different multithreaded programs, Paterson's algorithm, are executed on the SoC model in the \emph{SE} mode. The window size used is 12 for all gem5 traces, as the maximum length of a gem5 flow found in the official documentation is $12$.
The architectural slicing is based on {\tt packet id} and {\tt memory address} information of the packet. The traces are sliced based on the pair of packets and address information. Total-sliced refers to slicing by applying the windowing technique on architectural sliced traces. The number of non-zero edges drastically decreases in the Architectural and Total-sliced traces. The processed traces are $2$MB, $30$MB, and  $100$GB for {\sf Snoop}, {\sf Threads}, and {\sf Kernel} respectively.

\subsection{Interpreting the Results} 
After preparing the traces and collecting the message information, the model synthesis framework initiates the generation of solutions. This is achieved by iterative reduction of the model size generated by the model extraction Algorithm~\ref{algo:model-extract}. In our experiments, the algorithm terminates once it has found $200$ reduced models. These solutions encompass various models with different edge support that satisfy the constraints discussed in section~\ref{sec:model-gen}. To improve human comprehensibility, we carefully select $20$ solutions based on their size, giving preference to smaller solutions as they are deemed more understandable when converted to FSAs. The average acceptance ratio (AR) of these $20$ solutions is presented for both synthetic and gem5 traces, while the best AR among them is also reported. Additionally, the size of the model can serve as a valuable metric for evaluating the quality of a solution. 
We present the FSA model size for each type of trace, providing insights into the complexity of the synthesized models. Additionally, we report the accumulated run time (RT) required by our tool for different stages, such as causality graph building, graph annotation, solving consistency constraints, and evaluating the top $20$ solutions on the traces. These metrics are essential for evaluating the effectiveness and efficiency of our tool.

\subsection{Experiments with Synthetic Traces}

\begin{table*}[!ht]
    \centering
    \caption{Model synthesis results from Synthetic traces}
    \renewcommand{\arraystretch}{1.3}
    \resizebox{\textwidth}{!}{%
    \begin{tabular}{|c|c|c|c|c|c|c|c|c|c|c|c|c|c|c|c|c|c|c|c|}

    \hline
        \multicolumn{2}{ | c | }{{\multirow{3}{*}{}}} & \multicolumn{6}{c}{\centering {\sf Small-5} } & \multicolumn{6}{|c}{\centering {\sf Small-10}}& \multicolumn{6}{|c|}{\centering {\sf Small-20}}\\ \cline{3-20}
        \multicolumn{2}{ | c  }{} & \multicolumn{3}{|c}{\centering Unsliced} & \multicolumn{3}{|c}{\centering Window Sliced} & \multicolumn{3}{|c}{\centering Unsliced} & \multicolumn{3}{|c}{\centering Window Sliced} & \multicolumn{3}{|c}{\centering Unsliced} & \multicolumn{3}{|c|}{\centering Window Sliced}\\ \cline{3-20}
        \multicolumn{2}{ | c | }{} & Size & RT & AR & Size & RT & AR & Size & RT & AR & Size & RT & AR & Size & RT & AR & Size & RT & AR \\ \hline
        \multicolumn{2}{| c |}{\centering Average} & 54 & 31 & 54.19\% & 78 & 15 & 74.12\% & 58 & 33 & 63.97\% & 82 & 26 & 83.17\% & 55 & 31 & 65.50 \% & 79 & 19 & 79.56\% \\ \hline
        \multicolumn{2}{| c |}{\centering Best}    & 55 & 31 & 55.00   \% & 78 & 15 & 75.00\% & 59 & 33 & 64.89\% & 83 & 26 & 84.34\% & 55 & 31 & 70.91\% & 80 & 19 & 81.01\% \\ \hline \hline 
        
        \multicolumn{2}{ | c  }{} & \multicolumn{6}{|c}{\centering {\sf Large-5} } & \multicolumn{6}{|c}{\centering {\sf Large-10 }}& \multicolumn{6}{|c|}{\centering {\sf Large-20}}\\ \hline
        \multicolumn{2}{| c |}{\centering Average} & 89 & 22 & 59.38\% & 165 & 39 & 82.21\% & 88 & 26 & 54.05\% & 179 & 54 & 80.83\% & 86 & 30 & 60.86\% & 213 & 58 & 83.69\% \\ \hline
        \multicolumn{2}{| c |}{\centering Best}    & 89 & 22 & 60.50 \% & 166 & 39 & 83.07\% & 89 & 26 & 54.09\% & 181 & 54 & 82.04\% & 87 & 30 & 63.11\% & 215 & 58 & 85.57\% \\ \hline  

    \end{tabular}
    }
    \label{tab:synthetic_traces_results}
\end{table*}

\begin{table*}[!ht]
    \centering
    \caption{Model synthesis results from gem5 traces. The RT is in HH:MM:SS, hours, minutes, seconds format.}
    \renewcommand{\arraystretch}{1.3}
    \resizebox{\textwidth}{!}{%
    \begin{tabular}{|c|c|c|c|c|c|c|c|c|c|c|c|c|c|}

    \hline
        \multicolumn{14}{ | c | }{\centering {\sf Kernel}} \\ \hline 
        \multicolumn{2}{ | c | }{\multirow{2}{*}{}} & \multicolumn{3}{ p{3cm} |}{\centering Unsliced} & \multicolumn{3}{ p{3cm} |}{\centering Window Sliced} & \multicolumn{3}{ p{3cm} |}{\centering Architectural Sliced} & \multicolumn{3}{ p{3cm} |}{\centering Total Sliced} \\ \cline{3-14}
        \multicolumn{2}{ | c | }{} & Size & RT & AR & Size & RT & AR & Size & RT & AR & Size & RT & AR \\ \hline
        \multicolumn{2}{| c |}{\centering Average} & 209 & 7:42:03 & 71.06\% & 243 & 8:04:26 & 78.43\% & 140 & 25:14:38 & 79.00\% & 141 & 25:26:41 & 78.68\% \\ \hline
        \multicolumn{2}{| c |}{\centering Best}    & 214 & 7:42:03 & 71.07\% & 245 & 8:04:26 & 78.44\% & 141 & 25:14:38 & 79.03\% & 143 & 25:26:41 & 78.69\% \\ \hline \hline
        
        \multicolumn{14}{ | c | }{\centering {\sf Threads}} \\ \hline
        \multicolumn{2}{| c |}{\centering Average} & 216 & 9:05:00 & 97.39\% & 239 & 11:58:00 & 98.21\% & 122 & 2:45:18 & 98.16\% & 123 & 2:43:32 & 97.92\% \\ \hline
        \multicolumn{2}{| c |}{\centering Best}    & 216 & 9:05:00 & 97.40\% & 240 & 11:58:00 & 98.29\% & 122 & 2:45:18 & 98.21\% & 123 & 2:43:32 & 98.19\% \\ \hline \hline 
        
        \multicolumn{14}{ | c | }{\centering {\sf Snoop}} \\ \hline %
        \multicolumn{2}{| c |}{\centering Average} & 176 & 2:09:00 & 89.04\% & 179 & 4:14:00 & 95.72\% & 106 & 18:58:00 & 97.36\% & 101 & 18:32:00 & 97.37\% \\ \hline
        \multicolumn{2}{| c |}{\centering Best}    & 177 & 2:09:00 & 89.09\% & 181 & 4:14:00 & 96.01\% & 107 & 18:58:00 & 97.38\% & 104 & 18:32:00 & 97.38\% \\ \hline

    \end{tabular}
    }
    \label{gem5_traces_results}
\end{table*}

We run {\sf AutoModel} on the two sets of synthetic traces. Table~\ref{tab:synthetic_traces_results} reports size as model size, RT as run time in seconds, and AR as the acceptance ratio of our tool on those traces. When examining the size of the produced solution, readers should refer to the number of non-zero edges in the causality graphs built from these traces presented in the third and fourth columns of Table~\ref{tab:synt_analysis}. The largest solution size could be the number of non-zero edges of the causality graph of that trace. {\sf AutoModel} iteratively removes the highest number of those non-zero edges and checks if the graph is consistent with the node support. The result for each type of trace is divided into unsliced and sliced parts to highlight the benefit of applying slicing to the traces. Table~\ref{tab:synthetic_traces_results} shows that the model size in the small and large traces ranges from $54$ to $59$ and $86$ to $89$, respectively, for unsliced cases. This similarity in the model sizes reflects the similar number of non-zero edges of those traces as highlighted in Table~\ref{tab:synt_analysis}. 

It is interesting to note that the model size is smaller for both the {\sf Small-20} and {\sf Large-20} traces compared to {\sf Small-10} and {\sf Large-5}, respectively. This observation may seem counter-intuitive. However, it is important to consider that Algorithm~\ref{algo:reduce-model} iteratively reduces the set of initial models. We attribute these results to the stochastic nature of the SMT solver used for constraint solving. The solver generates random solutions for traces of the same type, leading to different constraint problems.
Consequently, the initial solutions generated for these constraint problems can vary. These differences in the initial sets of solutions can result in varying model sizes required to produce reduced models. It is worth noting that the model size, obtained through the invocation of procedures outlined in Algorithms~\ref{algo:model-extract} and ~\ref{algo:reduce-model}, can differ for traces of the same type for different executions.

The run time RT for small unsliced traces are mostly similar, as highlighted in Table~\ref{tab:synthetic_traces_results}. This phenomenon can be attributed to the fact that the major portion of the total runtime $(>80\%)$ is spent on constraint solving, which is dependent on the sizes of the causality graphs in terms of the numbers of edges with nonzero support~\cite{model_synth}. Table~\ref{tab:synt_analysis} shows that all small traces have $214$ nonzero edges. Therefore, solving constraint problems generated from similar causality graphs incurs similar complexity,
and its runtime performance is largely independent of trace lengths. This observation contrasts our method against the previous work~\cite{natasa2020}. However, {\sf Small-10} takes two seconds more, contributed by the evaluation time for a larger model, as we see this trace generates the largest model compared to other small traces.

We see a gradual increase in RT for unsliced large traces with the increase of nonzero edges in the causality graph. However, the {\sf Large-20} has a relatively higher RT, which could be attributed to the comparatively longer trace evaluation time even though the generated model size is smaller. We also see a tendency that a smaller model generation usually takes a longer time, even though the trace and causality graph remains similar. The root of this phenomenon lies in the iterative reduction of the model sizes. If a smaller solution is achievable, Algorithms~\ref{algo:model-extract} and \ref{algo:reduce-model} will spend more time to reach that solution.
In contrast, if a smaller solution is not achievable, the tool will report that solution immediately, saving some run time.  Interestingly, we observe that the RT for unsliced small traces is longer than for unsliced large traces. This phenomenon can be attributed to the time spent in Algorithm~\ref{algo:reduce-model}. Notably, we observe that Algorithm~\ref{algo:reduce-model} takes more time to solve the constraints in the small traces than the large traces.

Table~\ref{tab:synthetic_traces_results} also presents the AR for the synthetic traces. Generally, AR increases as the model size increases for a given trace. However, it is worth noting that a solution with perfect AR but a significantly large size may not be desirable, as it may not accurately capture the aggregate behavior in the traces. It is also important to mention that in our algorithm, the AR is not used in the model generation process, and therefore the higher AR is not prioritized. Finding the optimal balance between model size and AR would be an interesting research direction, but it falls beyond the scope of this work. The AR values for different traces indicate the quality of the model for each specific trace, but they are unrelated to each other. Additionally, evaluation has its own challenges, as discussed in Section~\ref{sec:eval}. Therefore, readers should compare ARs between optimization techniques applied to the same trace.


\subsubsection{Mining from multiple-traces}
As described in section \ref{subsec:multiple}, {\sf AutoModel} can handle multiple traces for extracting the execution model. Table~\ref{tab:multiple_res} reports models synthesis from multiple traces. The features of the multiple traces are described in Table~\ref{tab:synt_analysis}. Our tool can reduce the number of non-zero edges to a couple of folds. An interesting observation is that solving constraints for this large trace takes only $132$ seconds in unsliced cases. This shorter synthesis time is one of the strengths of this tool compared to existing methods in this domain, as they usually suffer from exponential run time as the trace length increases. We see the AR increases in the windowing showing the benefit of windowing on multiple traces too. 

\begin{table}[!ht]
\centering
\caption{Model Synthesis results from {\sf Multiple} synthetic traces}
\label{tab:multiple_res}
\renewcommand{\arraystretch}{1.3}
\begin{tabular}{|c|ccc|ccc|}
\hline
\multirow{2}{*}{} & \multicolumn{3}{c|}{Unsliced} & \multicolumn{3}{c|}{Window sliced} \\ \cline{2-7} 
        & \multicolumn{1}{l|}{Size} & \multicolumn{1}{l|}{RT} & AR      & \multicolumn{1}{l|}{Size} & \multicolumn{1}{l|}{RT} & AR      \\ \hline
Average & \multicolumn{1}{l|}{69}  & \multicolumn{1}{l|}{132} & 47.70\% & \multicolumn{1}{l|}{70}  & \multicolumn{1}{l|}{64} & 84.78\% \\ \hline
Best    & \multicolumn{1}{l|}{70}  & \multicolumn{1}{l|}{130} & 54.17\% & \multicolumn{1}{l|}{72}  & \multicolumn{1}{l|}{68} & 87.17\% \\ \hline
\end{tabular}
\end{table}

\subsubsection{Discussion on windowing}

The obtained models from the window-sliced synthetic traces demonstrate improvements in AR for all six synthetic traces. The synthetic traces are generated with controlled interleaving, ensuring that there are no more than $10$ messages between two messages from the same branch of a flow. This control of interleaving enables us to use a window size of $10$ for slicing the synthetic traces, resulting in the window-sliced traces of the corresponding (unsliced) traces.

\begin{figure}[!ht]
    \centering
    \includegraphics[width = 0.9\columnwidth]{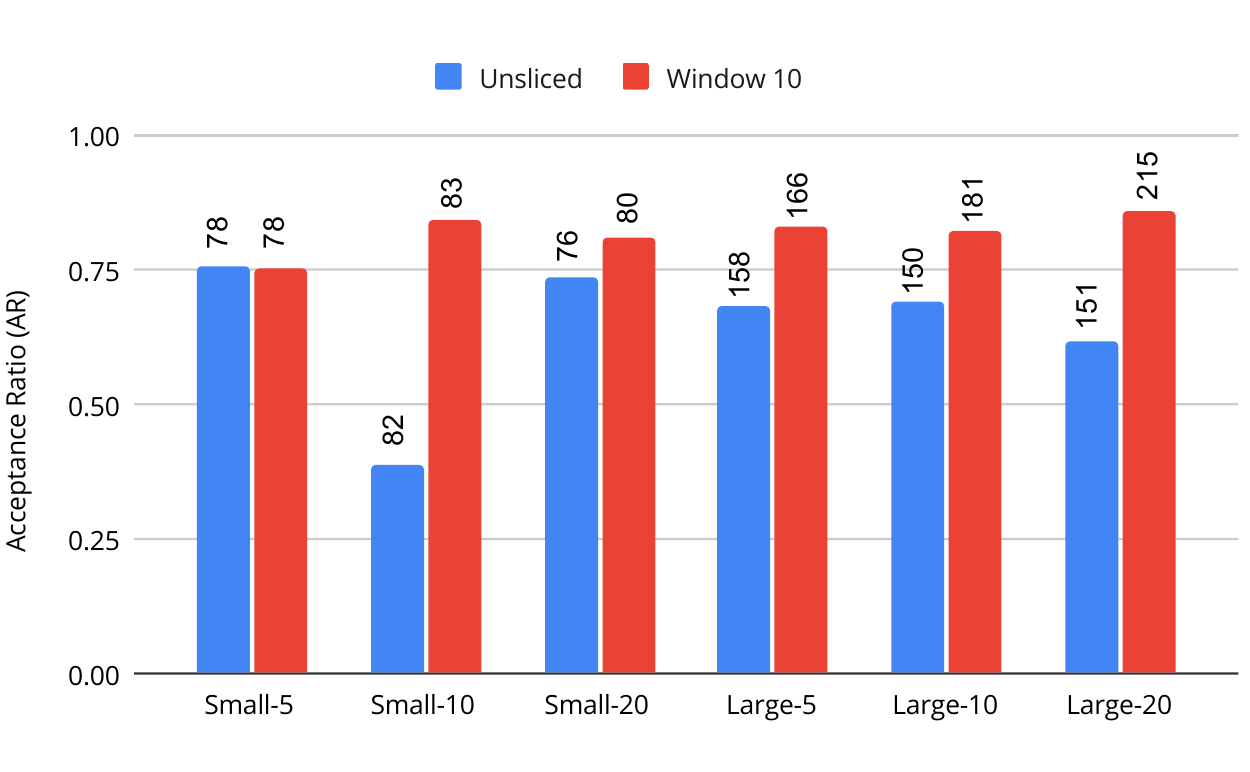}
    \caption{Best acceptance ratio comparison between unsliced and window-sliced synthetic traces. Similar-sized (the number on the bar head) models from the sliced traces have better AR.}
    \label{fig:ResultsSyntheticABest}
\end{figure}

Table~\ref{tab:synthetic_traces_results} illustrates that windowing enhances the AR by a minimum of $14\%$ and a maximum of $36\%$ for the {\sf Small-20} and {\sf Small-5} traces, respectively. Additionally, the windowing approach increases model size by $41\%$ and $45\%$ in these experiments. Notably, we observe further improvements in AR for the large traces. Windowing increases AR by a minimum of $37\%$ and a maximum of $50\%$ in {\sf Large-5} and {\sf Large-10} traces, respectively. However, it is important to note that this increment in AR may be influenced by the increase in model size, which is $86\%$ and $103\%$ for the respective cases. However, windowing offers better AR for similar model sizes than the unsliced traces. Fig.~\ref{fig:ResultsSyntheticABest} reports that windowing has higher AR for a comparable model size from the unsliced trace. It should be noted that RT for the models from the unsliced traces in Fig.~\ref{fig:ResultsSyntheticABest} is not reported as these models were not reported in the first $200$ iterations in the unsliced traces. Therefore, windowing improves the quality of synthesized models and helps extract those models faster. 

A consistent trend across the experiments indicates that AR generally increases as the traces become longer. Low AR could also be a consequence of a high degree of interleaving in the traces. The key advantage of windowing is its ability to reduce approximately three-quarters of non-zero edges in the small traces and around two-thirds of non-zero edges in the large traces in the synthesized models. 

Table~\ref{tab:synthetic_traces_results} also presents the runtime (RT) for each window-sliced trace. The unsliced small traces exhibit longer RT than their window-sliced counterparts, while the opposite is true for the large traces. The shorter RT in window-sliced small traces than the unsliced is because unsliced reduces the model further than the window-sliced, resulting in longer RT. However, for the large traces, window-sliced have increased RT attributed to the considerably bigger size of the causality graph of this trace. Annotating a large graph takes longer than a shorter one, and bigger solutions will take longer to evaluate. Both are true for large traces. 

The synthesized model size is nearly doubled in the sliced than the unsliced large traces. The increase in model size for windowing is because windowing reduced the number of non-zero edges in the causality graph, as shown in Table~\ref{tab:synt_analysis}. In this case, the solver has less number of branches to solve the constraints problem. In other words, the increase in the model size indicates fewer inconsistent or false edges are available when windowing is applied. We also see that windowing hasn't reduced the non-zero edges in small traces as much as in large traces. Therefore, the reduction of RT in window-sliced small traces and the increase in RT for window-sliced large traces are justified.
In summary, the results in Table~\ref{tab:synthetic_traces_results} highlight the benefits of using window-slicing techniques, which lead to improved acceptance ratios and reduced non-zero edges in the synthesized models. Moreover, the RT shed light on the trade-off between model size, interleaving, and the efficiency of the synthesis process. In this experiment, we fix the size of the sliding window $w$ in \emph{Trace2Model} to $3$ as suggested in~\cite{natasa2020}.
\emph{Trace2Model} suffers from a considerably long run time when applied to synthetic traces. It does not complete even for the {\sf small} traces after $30$ minutes.  
Therefore, no results with \emph{Trace2Model} are reported.

\begin{figure}[!ht]
    \centering
    \includegraphics[width = 0.9\columnwidth]{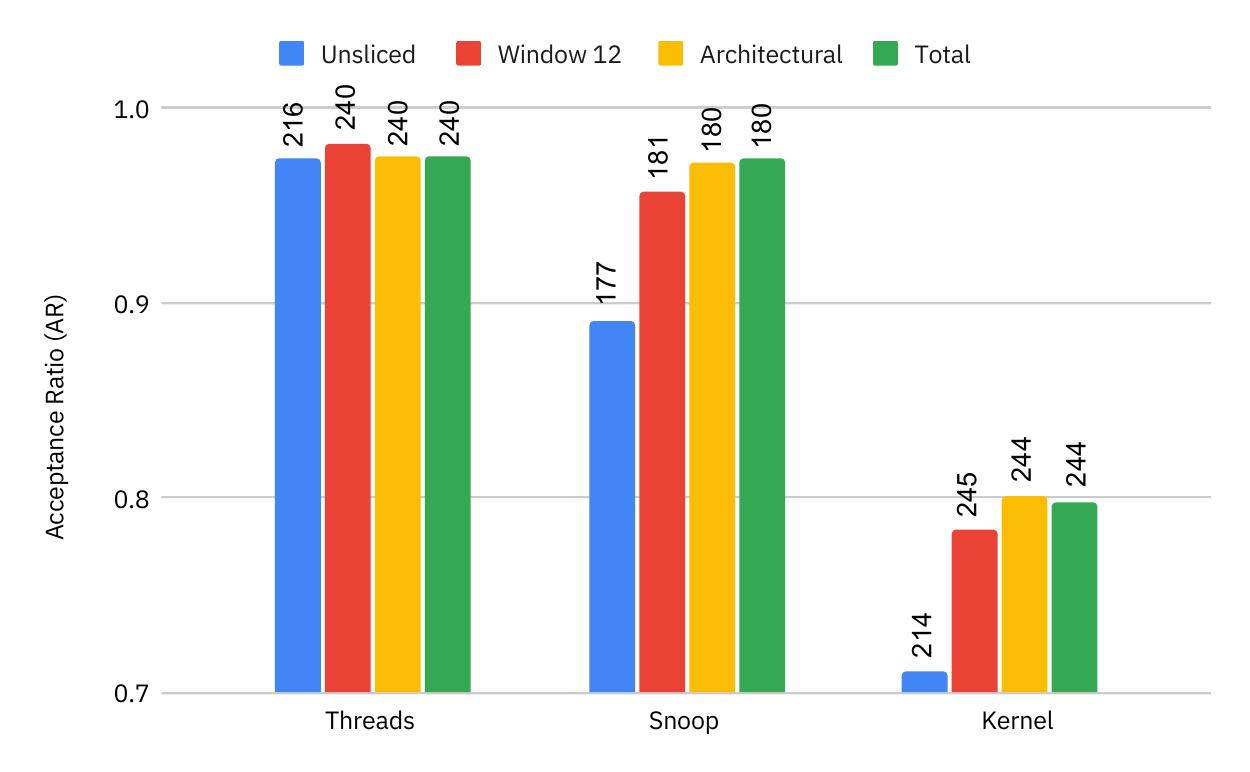}
    \caption{Best acceptance ratio comparison between unsliced and window-sliced gem5 traces.}
    \label{fig:ResultsGem5ABest}
\end{figure}


\subsection{Experiments with gem5 Traces}
The evaluation of gem5 traces involves two aspects. The first aspect examines the results using unsliced traces, while the second aspect explores the benefits of slicing the gem5 traces to improve the model synthesis.

In the case of unsliced traces, there is a noticeable reduction of non-zero edges to the synthesized models, a ratio of approximately 10:1 for the {\sf Kernel} trace. The reduction is even more significant for the {\sf Threads} traces. It may seem counterintuitive that the RT for the {\sf Threads} traces is higher than that of the {\sf Kernel} trace, despite the latter being significantly longer. However, this phenomenon can be explained by the slightly larger size and higher number of non-zero edges in the causality graph of the {\sf Threads} trace. The increased complexity of the causality graph and the constraints contributes to the longer RT.
Furthermore, the {\sf Kernel} trace includes numerous boot sequence messages that facilitate coordinating tasks such as loading the BIOS, detecting and initializing hardware devices, and ensuring a successful OS boot. These messages play a crucial role in the early stages of the Linux boot process, where the BIOS or UEFI firmware initializes the hardware, and the Linux kernel takes control to continue the boot procedure. This trace's high degree of isolated messages leads to a shorter AR. Considering these factors, the longer RT for the {\sf Kernel} traces is acceptable. Comparing the results from the {\sf Threads}, {\sf Kernel}, and {\sf Snoop} traces, the size and RT for the {\sf Snoop} trace becomes intuitive, similar to the {\sf Threads} and {\sf Kernel} traces. Notably, the {\sf Threads} trace demonstrates the highest AR among the unsliced gem5 traces.

\subsubsection{Discussion on slicing}
Similar to the synthetic traces, we observed improvements in model size and acceptance ratios in the gem5 traces after applying different slicing techniques. We used a window size of twelve for window slicing on all three gem5 traces.
On the {\sf Kernel} trace, windowing increased the AR by approximately $10\%$ while enlarging the model size by more than $14\%$. However, architectural slicing yielded even more significant improvements, enhancing AR further while reducing the model size by over $42\%$. For the {\sf Snoop} trace, architectural slicing improved AR by more than $9\%$ and reduced the model size by about $39\%$. All Sliced traces showed further reductions in model size. However, windowing only slightly improved AR on the {\sf Threads} trace but led to an approximately $11\%$ increase in model size. In contrast, architectural slicing achieved a model size reduction of more than $43\%$ for the {\sf Threads} trace while maintaining AR close to that of windowing. Figure~\ref{fig:ResultsGem5ABest} highlights that we can maintain the same or better AR with smaller models by applying different slicing techniques.

The findings indicate that architectural slicing is better suited for gem5 traces than windowing. This is crucial to note that in realistic scenarios, numerous architectural information exists while the SoC design is in the RTL stage. Therefore incorporating the required structural information can benefit the model synthesis to a great extent. However, the same is not true for post-silicon validation, while slicing may be challenging due to limited observability or the proprietary nature of commercial SoCs~\cite{observability}. Therefore, some preprocessing or repair of the post-silicon traces will be required before using this method in post-silicon debugging.

\begin{figure*}[!ht]
\begin{tabular}{cc}
    \includegraphics[height=1.4in]{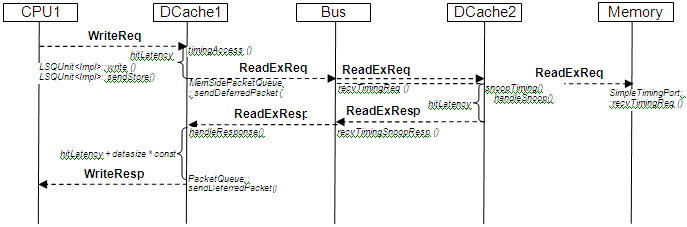}
    & 
    \includegraphics[height=1.4in]{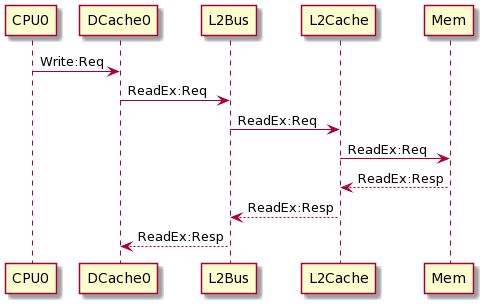}\\
    (a) & (b)
\end{tabular}
    \caption{(a) A memory write protocol specification in gem5 documentation showing the case of write miss in DCache; (b) A path in the extracted FSA model in a sequence diagram showing the memory write operations executed by the system simulation model.}
    \label{fig:gem5-spec}
\end{figure*}

\begin{figure*}[!ht]
\begin{tabular}{cc}
    \includegraphics[height=1.4in]{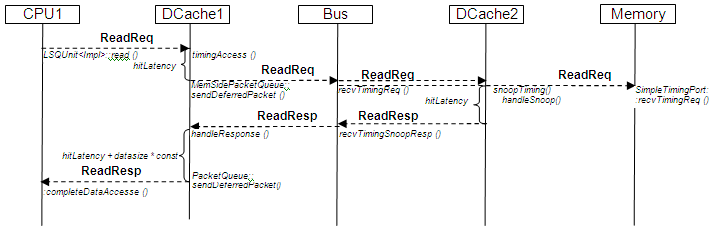}
    & 
    \includegraphics[height=1.4in, width=2.3in]{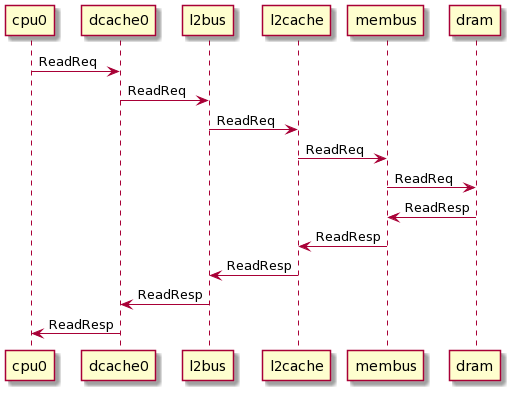}\\
    (a) & (b)
\end{tabular}
    \caption{(a) A memory read protocol specification showing the case of read miss in DCache; (b) A path extracted using the proposed method in a sequence diagram showing the memory read miss operation executed by the gem5 simulation model.}
    \label{fig:gem5_readreq}
\end{figure*}

\subsection{Comparison with other tools}
Our research is motivated by the limitations of existing tools in effectively extracting meaningful patterns from complex SoC execution traces. Traditional FSA extraction methods like \emph{Trace2Model} have shown ineffectiveness in extracting models from lengthy gem5 traces spanning 48 hours. While sequential pattern mining tools like \emph{Perracotta} can efficiently identify shorter sequential patterns, these patterns provide limited insights into execution traces or scenarios. On the other hand, the tool \emph{Synoptic} captures all temporal causality but lacks mechanisms to remove false-related messages, leading to excessively large solution sizes that are challenging to comprehend.

We compare our method with \emph{Trace2Model} on synthetic traces too. We selected \emph{Trace2Model} and our method as both attempts to synthesize FSA models to accept input traces using constraint-solving techniques. Other works~\cite{Yang:2006, Liu:2013} focus solely on finding interesting sequential patterns from traces and thus are not included in the comparison. As discussed in Section~\ref{sec:gap}, \emph{Trace2Model} does not consider the concurrency inherent in the traces, unlike our method, and relies on identifying temporal dependencies among messages, resulting in extracted models that may contain message sequences that lack coherence.


\begin{figure}[!ht]
\centering
\includegraphics[height=1.2in]{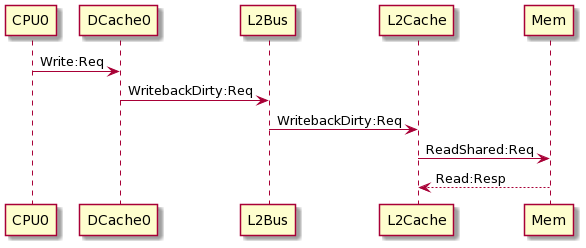}
\caption{Another example of message sequences for dirty memory block write-back operations from the extracted FSA model.}
\label{fig:gem5_dirtywb}
\end{figure}

\subsection{Case Study: Using Model Synthesis for Debugging gem5 Message Flows} 
The official documents of gem5 do not fully specify all the communication protocols used to implement different system-level functions. As a result, the exact size of the ground truth model remains unknown. However, we extract specific sections of the FSAs constructed from gem5 traces. These extracted parts offer valuable insights into the underlying implementation of various system-level protocols. While the complete specifications are unavailable, the extracted FSAs provide useful information to enhance our understanding of the system's communication mechanisms.

Fig.~\ref{fig:gem5-spec}(b) shows an example message flow in the FSA model extracted from the trace, while Fig.~\ref{fig:gem5-spec}(a) shows the similar memory write protocol from the gem5 system documentation~\cite{gem5:protocol:write}.  
It is unclear whether the protocol in Fig.~\ref{fig:gem5-spec}(a) is used in the gem5 model. 
Comparing Fig.~\ref{fig:gem5-spec}(b) to (a), the protocol in the model extracted using our method is different from the protocol shown in Fig.~\ref{fig:gem5-spec}(a) in two aspects. 
\begin{enumerate}
    \item Our extracted protocol includes a response message returned from {\tt Mem} to {\tt L2Cache} while this message is missing from Fig.~\ref{fig:gem5-spec}(a).  
    We believe that missing messages should be included in the model as a memory block must be fetched from {\tt Mem} to {\tt L2Cache} and then to {\tt DCache0} in case of a write miss. 
    \item Our extracted protocol does not have the write response message from {\tt DCache0} to {\tt CPU0} as in Fig.~\ref{fig:gem5-spec}(a).  
    We think the missing message could be redundant and not be used in gem5 as it is typical that a write operation from the CPU completes at the cache without the need to return a response.
\end{enumerate} 

We also analyze a ReadMiss scenario for {\tt CPU0} using message sequence charts (MSC) as presented in the official documentation of gem5, shown in Fig.~\ref{fig:gem5_readreq}(a). Upon careful examination, we find that the official MSC does not explain how the parallel ReadReqs originating from the {\tt Bus} are handled. Model synthesis can help fill this gap. We extract a corresponding MSC from the {\sf Snoop} traces, as depicted in Fig.~\ref{fig:gem5_readreq}(b), which sheds light on how the ReadReqs are handled in the {\tt membus} and {\tt dram}. This example shows that specifications may become out-of-date or deviate from actual implementation. 
Another example flow in the extracted model is shown in Fig.~\ref{fig:gem5_dirtywb}. It shows a memory write protocol where a dirty block is written back to the L2 cache under an L1 miss. 
In this example, we can find that the simulation model utilizes a write-allocate policy on misses, despite some response messages being cut off. However, the official documents lack this specific write protocol specification. By leveraging the synthesis and analysis of concise models from execution traces, we can address the specification problem.
From the three case studies conducted on gem5 traces, below are summaries of some important observations of the proposed mining method. 
\begin{itemize}
\item It can accurately extract message flows from complex and very long SoC traces.
\item It can mine message flows missing from the official documentation, \eg~Fig.~\ref{fig:gem5_dirtywb}.
\item It can mine message flows that are not accurately specified in the official documentation, \eg~Fig.~\ref{fig:gem5-spec} and Fig.~\ref{fig:gem5_readreq}.
\end{itemize}

\section{Conclusion}
\label{sec:conclusion}
Our method addresses the challenging task of modeling highly concurrent SoC communication traces obtained from complex SoC designs capable of running operating systems. We have demonstrated the efficiency of our approach in inferring reduced and abstract models from very long traces. These extracted models offer valuable insights into system-level protocols, facilitating a better understanding of how different components coordinate operations within the system. We have presented three algorithms for extracting models from the traces and one algorithm for evaluating the extracted models, discussing the associated challenges in both processes. A comprehensive case study on the popular gem5 tool is provided to validate our approach.

Moving forward, we aim to enhance our method by incorporating user insights, allowing users to guide the causality graph-building process effectively. By providing users with greater control, we anticipate improving the accuracy and usefulness of the synthesized models. In general, our work significantly contributes to the field of model synthesis for complex system designs, opening new possibilities for analyzing and comprehending highly concurrent communication traces.







\bibliographystyle{unsrt}
\bibliography{bibfiles/references,bibfiles/soc,bibfiles/others,bibfiles/dm,ref}

\end{document}